%% file: h1draft2.tex
\documentclass[12pt]{article}

\usepackage[dvips]{changebar}
\usepackage{ccaption}
\usepackage{epsfig}
\usepackage{amsmath}
\usepackage{hhline}
\usepackage{amssymb}
\usepackage{times}
\usepackage[varg]{txfonts}
\usepackage{cite}
\usepackage{dcolumn}
\newcolumntype{d}[1]{D{.}{.}{#1}}

\newlength{\dinwidth}
\newlength{\dinmargin}
\begin{document}  

\newcommand{\pom}{{I\!\!P}}
\newcommand{\reg}{{I\!\!R}}
\newcommand{\slowpi}{\pi_{\mathit{slow}}}
\newcommand{\fiidiii}{F_2^{D(3)}}
\newcommand{\fiidiiiarg}{\fiidiii\,(\beta,\,Q^2,\,x)}
\newcommand{\n}{1.19\pm 0.06 (stat.) \pm0.07 (syst.)}
\newcommand{\nz}{1.30\pm 0.08 (stat.)^{+0.08}_{-0.14} (syst.)}
\newcommand{\fiidiiiful}{F_2^{D(4)}\,(\beta,\,Q^2,\,x,\,t)}
\newcommand{\fiipom}{\tilde F_2^D}
\newcommand{\ALPHA}{1.10\pm0.03 (stat.) \pm0.04 (syst.)}
\newcommand{\ALPHAZ}{1.15\pm0.04 (stat.)^{+0.04}_{-0.07} (syst.)}
\newcommand{\fiipomarg}{\fiipom\,(\beta,\,Q^2)}
\newcommand{\pomflux}{f_{\pom / p}}
\newcommand{\nxpom}{1.19\pm 0.06 (stat.) \pm0.07 (syst.)}
\newcommand {\gapprox}
   {\raisebox{-0.7ex}{$\stackrel {\textstyle>}{\sim}$}}
\newcommand {\lapprox}
   {\raisebox{-0.7ex}{$\stackrel {\textstyle<}{\sim}$}}
\def\gsim{\,\lower.25ex\hbox{$\scriptstyle\sim$}\kern-1.30ex%
\raise 0.55ex\hbox{$\scriptstyle >$}\,}
\def\lsim{\,\lower.25ex\hbox{$\scriptstyle\sim$}\kern-1.30ex%
\raise 0.55ex\hbox{$\scriptstyle <$}\,}
\newcommand{\pomfluxarg}{f_{\pom / p}\,(x_\pom)}
\newcommand{\dsf}{\mbox{$F_2^{D(3)}$}}
\newcommand{\dsfva}{\mbox{$F_2^{D(3)}(\beta,Q^2,x_{I\!\!P})$}}
\newcommand{\dsfvb}{\mbox{$F_2^{D(3)}(\beta,Q^2,x)$}}
\newcommand{\dsfpom}{$F_2^{I\!\!P}$}
\newcommand{\gap}{\stackrel{>}{\sim}}
\newcommand{\lap}{\stackrel{<}{\sim}}
\newcommand{\fem}{$F_2^{em}$}
\newcommand{\tsnmp}{$\tilde{\sigma}_{NC}(e^{\mp})$}
\newcommand{\tsnm}{$\tilde{\sigma}_{NC}(e^-)$}
\newcommand{\tsnp}{$\tilde{\sigma}_{NC}(e^+)$}
\newcommand{\st}{$\star$}
\newcommand{\sst}{$\star \star$}
\newcommand{\ssst}{$\star \star \star$}
\newcommand{\sssst}{$\star \star \star \star$}
\newcommand{\tw}{\theta_W}
\newcommand{\sw}{\sin{\theta_W}}
\newcommand{\cw}{\cos{\theta_W}}
\newcommand{\sww}{\sin^2{\theta_W}}
\newcommand{\cww}{\cos^2{\theta_W}}
\newcommand{\trm}{m_{\perp}}
\newcommand{\trp}{p_{\perp}}
\newcommand{\trmm}{m_{\perp}^2}
\newcommand{\trpp}{p_{\perp}^2}
\newcommand{\alp}{\alpha_s}
\newcommand{\alps}{$\alpha_s$}
\newcommand{\sqrts}{$\sqrt{s}$}
\newcommand{\LO}{$O(\alpha_s^0)$}
\newcommand{\Oa}{$O(\alpha_s)$}
\newcommand{\Oaa}{$O(\alpha_s^2)$}
\newcommand{\PT}{p_{\perp}}
\newcommand{\JPSI}{J/\psi}
\newcommand{\sh}{\hat{s}}
\newcommand{\uh}{\hat{u}}
\newcommand{\MP}{m_{J/\psi}}
\newcommand{\PO}{I\!\!P}
\newcommand{\xbj}{x}

\newcommand{\myx}{\ensuremath{x} }
\newcommand{\kperp}{\ensuremath{k_{\perp}^2} }
\newcommand{\deta}{\ensuremath{|\Delta\eta^*|} }
\newcommand{\mrsq}{\ensuremath{\mu^2_{r}} }

\newcommand{\xpom}{x_{\PO}}
\newcommand{\ttbs}{\char'134}
\newcommand{\xpomlo}{3\times10^{-4}}  
\newcommand{\xpomup}{0.05}  
\newcommand{\dgr}{^\circ}
\newcommand{\pbarnt}{\,\mbox{{\rm pb$^{-1}$}}}
\newcommand{\gev}{\,\mbox{GeV}}
\newcommand{\WBoson}{\mbox{$W$}}
\newcommand{\fbarn}{\,\mbox{{\rm fb}}}
\newcommand{\fbarnt}{\,\mbox{{\rm fb$^{-1}$}}}

\newcommand{\qsq}{\ensuremath{Q^2} }
\newcommand{\gevsq}{\ensuremath{\mathrm{GeV}^2} }
\newcommand{\et}{\ensuremath{E_t^*} }
\newcommand{\esq}{\ensuremath{E_t^2} }
\newcommand{\esqbar}{\ensuremath{\bar{E}_t^{*2}} }
\newcommand{\rap}{\ensuremath{\eta^*} }
\newcommand{\gp}{\ensuremath{\gamma^*}p }
\newcommand{\dsiget}{\ensuremath{{\rm d}\sigma_{ep}/{\rm d}E_t^*} }
\newcommand{\dsigrap}{\ensuremath{{\rm d}\sigma_{ep}/{\rm d}\eta^*} }

\def\Journal#1#2#3#4{{#1} {\bf #2} (#3) #4}
\def\NCA{\em Nuovo Cimento}
\def\NIM{\em Nucl. Instrum. Methods}
\def\NIMA{{\em Nucl. Inst. Meth.} {\bf A}}
\def\NPB{{\em Nucl. Phys.}   {\bf B}}
\def\PLB{{\em Phys. Lett.}   {\bf B}}
\def\PRL{{\em Phys. Rev. Lett.}}
\def\PRD{{\em Phys. Rev.}    {\bf D}}
\def\ZPC{{\em Z. Phys.}      {\bf C}}
\def\EJC{{\em Eur. Phys. J.} {\bf C}}
\def\CPC{{\em Comp. Phys. Commun.}}

\begin{titlepage}

\begin{figure}[!t]
DESY--03--160 \hfill ISSN 0418--9833\\ 
October 2003 
\end{figure}
\bigskip


\vspace*{2cm}

\begin{center}
\begin{Large}
  {\bf 
     Inclusive Dijet Production at Low Bjorken-\boldmath{\em x} \\ 
     in Deep Inelastic Scattering
  } 
       
\vspace{2cm}

H1 Collaboration

\end{Large}
\end{center}

\vspace{2cm}

\begin{abstract}
    
Dijet production in deep inelastic $ep$ scattering is investigated in the region
of low values of the Bjorken-variable~$x$ \mbox{($10^{-4} < x < 10^{-2}$)} and low
photon virtualities~$Q^2$ ($5 < Q^2 < 100$~GeV$^2$).  The measured dijet
cross sections are compared with perturbative QCD calculations in
next-to-leading order.  For most dijet variables studied, these calculations can
provide a reasonable description of the data over the full phase space region
covered, including the region of very low $x$.  However, large discrepancies are
observed for events with small separation in azimuth between the two
highest transverse momentum
jets. This region of phase space is described better by predictions based on the CCFM
evolution equation, which incorporates $k_{t}$ factorized unintegrated parton
distributions. A reasonable description is also obtained using the Color Dipole Model or
models incorporating virtual photon structure.

\end{abstract}

\vspace{1.5cm}

\begin{center}
To be submitted to Eur.\,Phys.\,J.\,C
\end{center}

\end{titlepage}

 \begin{flushleft}
   \input{h1auts}

 \end{flushleft}

\newpage
\section{Introduction}
\noindent

\begin{figure}[t]
\center
\epsfig{file=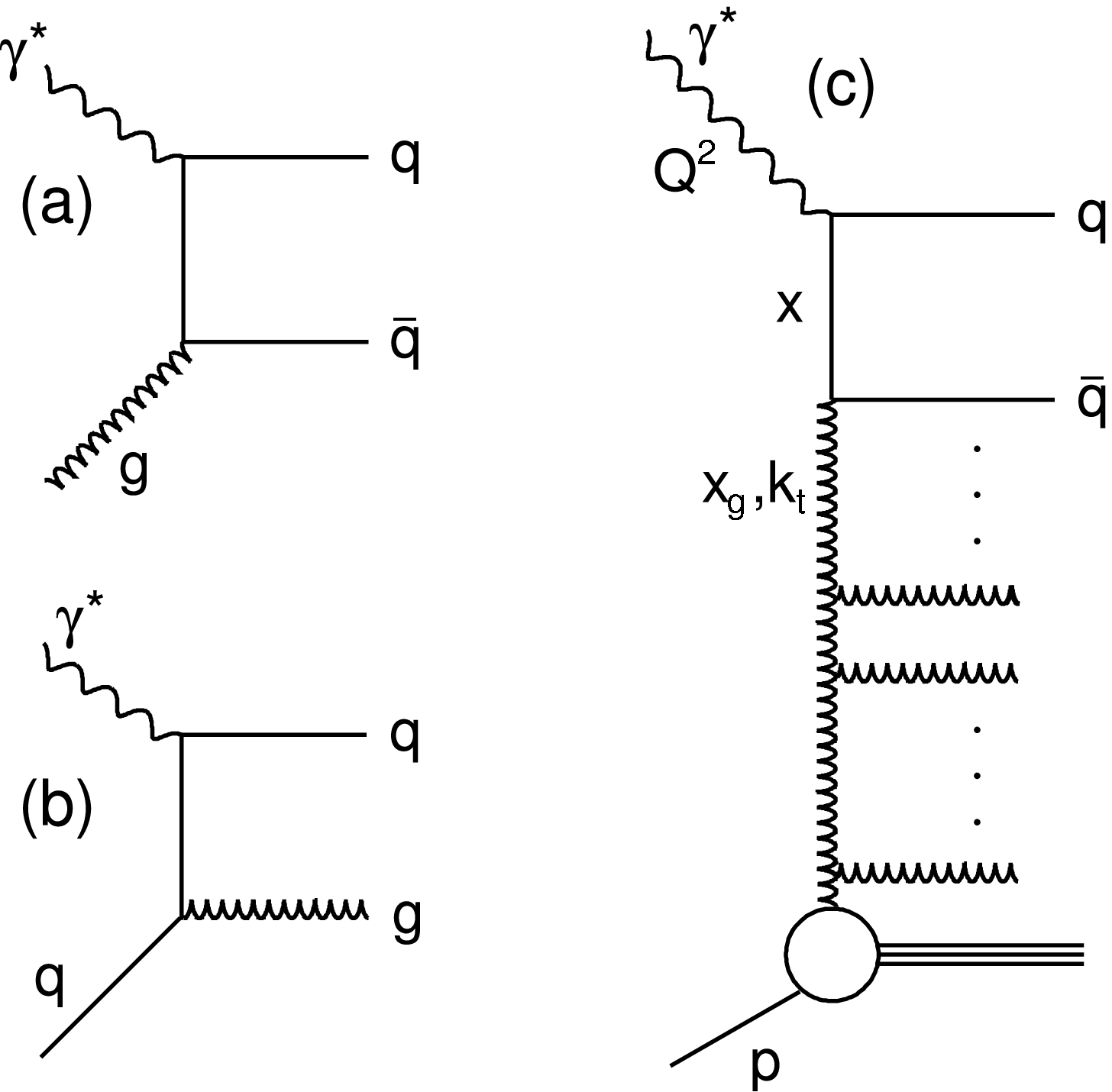,width=0.8\textwidth}
\caption{
Leading order diagrams for dijet production in $ep$ scattering. (a)
photon-gluon fusion and (b) QCD-Compton process. (c) parton cascade diagram:
$k_t$ denotes the transverse momenta of the exchanged gluons, $x_g$ 
the fractional longitudinal momentum of the gluon taking part in the hard 
process and $x$ is the Bjorken scaling variable. }
\label{fig:lo} 
\end{figure}

Dijet production in deep inelastic lepton-proton scattering (DIS) provides an
important testing ground for Quantum Chromodynamics (QCD). At HERA,
data are collected over a large range of the negative
four-momentum transfer squared,
$Q^{2}$, the Bjorken-variable $x$ and the transverse energy, $E_{T}$, of the
observed jets. HERA dijet
data may be used to gain insight into the dynamics of the parton cascade
exchanged in low-$x$ lepton-proton interactions.  Since in this
region of phase space photon-gluon fusion (Figure\,\ref{fig:lo}a) is the dominant
underlying process for dijet production, such measurements open the
possibility of studying the unintegrated gluon
distribution, first introduced in~\cite{FKL,FKL1}.

In leading order (LO), {\it i.e.} \Oa, dijet production in DIS is described by the
boson-gluon fusion and QCD-Compton processes (Figure\,\ref{fig:lo}a, b).  The cross
section depends on the fractional momentum $x$ of the incoming parton, where
the probability distribution of $x$ is given by the parton density functions
(PDFs) of the proton.  The evolution of the PDFs with the factorization scale,
$\mu_f^2$, is generally described by the DGLAP equations~\cite{DGLAP}.  To
leading logarithmic accuracy, this is equivalent to the exchange of a parton
cascade, with the exchanged partons strongly ordered in 
virtuality up to $Q^2$. For low $x$ this becomes approximately an ordering in 
$k_t$, the transverse momentum of the partons in the cascade (Fig.~\ref{fig:lo}c).
This paradigm has been highly successful in the
description of jet production at HERA at large values of $Q^2$ or
$E_{T}^2$ of the jet~\cite{H12JETS,H13JETS,ZEUS2JETS,ZEUSALPHAS}.

The DGLAP approximation is expected to breakdown at low $x$, as it
only resums leading logarithms in $Q^{2}$ and neglects contributions from $\log
1/x$ terms, which are present in the full perturbative expansion.  This
breakdown may have been observed in forward jet and forward particle
production at HERA~\cite{JGCN, ZEUS1, ZEUS2, h1incl}.

Several theoretical approaches exist which account for low-$x$ effects not
incorporated into the standard DGLAP approach.  At very low values of $x$ it is
believed that the theoretically most appropriate description is given by the BFKL
evolution equations~\cite{FKL1, BL}, which resum large logarithms of $1/x$ up to
all orders.  The BFKL resummation imposes no restriction on the
ordering of the transverse momenta within the parton cascade. Thus off-shell matrix
elements have to be used together with an unintegrated gluon distribution
function, $f(x,\tilde{\mu}_f^2,k_{t})$, which depends on the gluon transverse
momentum $k_t$ as well as $x$ and a hard scale $\tilde{\mu}_f$.  
A promising approach to parton evolution at low and larger values of $x$ is
given by the CCFM~\cite{CCFM} evolution equation, which, by means of
angular-ordered parton emission, is equivalent to the BFKL ansatz for $x
\rightarrow 0$, while reproducing the DGLAP equations at large $x$.

Experimentally, deviations from the DGLAP approach may best be
observed by selecting events in a phase space region where the main
assumption, the strong ordering in $k_t$ of the exchanged partons
in the cascade, is no
longer expected to be fulfilled. This is the case at low $x$. Parton emission along
the exchanged gluon ladder (Fig.~\ref{fig:lo}c) increases with decreasing $x$.
This may lead to large transverse momenta of the partons entering the hard
scattering process, such that in the hadronic~(photon-proton) center-of-mass system (cms) 
the two partons produced in the hard
scattering process (Fig~\ref{fig:lo}a,b) are no longer balanced in transverse
momentum. For the final state studied here, one then expects an excess of
events in which the two hardest jets are no longer back-to-back in azimuth.
Such configurations are not included in DGLAP-based
calculations, necessitating the inclusion of additional contributions when
calculating cross sections at low $x$.

An alternative approach to modelling additional contributions~\cite{JUJOKU}
due to non-$k_t$-ordered parton cascades is given by the concept of virtual
photon structure.  This approach mimics higher order QCD effects at low $x$ by
introducing a second $k_t$-ordered parton cascade on the resolved photon side,
evolving according to the DGLAP formalism. This
resolved contribution is expected to contribute for squared transverse jet
energies, $E^{2}_T$, greater than $Q^{2}$, which is the case for most of the phase
space of the present analysis.  Virtual photon structure is expected to be
suppressed with increasing $Q^2$.  
Leading-order QCD models which include the
effects of a resolved component to the virtual photon have been successful in
describing dijet production at low $Q^2$~\cite{H1DIJETSRES1}.

The aim of this paper is to provide new data in order to identify those regions of
phase space in which next-to-leading order (NLO) DGLAP-based QCD calculations
are able to correctly describe the underlying dynamics of the exchanged parton
cascade and those in which the measurements deviate from the DGLAP based predictions. 
Where deviations are observed, comparisons of the data with other QCD models are
performed.  Dijet production at low $x$ and low $Q^{2}$ is an appropriate tool
for this purpose as the jet topology reflects the dynamics of the
parton cascade~\cite{SZCU,ASKE,FORS}.  Therefore,
dijet cross sections are measured multi-differentially as a function of
observables particularly sensitive to low-$x$ dynamics, considerably extending
an earlier analysis~\cite{r2js} in terms of the observables studied,
the kinematic reach and statistical precision.

\section{Experimental Environment}

The measurement presented is based on data collected with the H1
detector at HERA during the years 1996 and 1997.  During this
period the HERA collider was operated with positrons\footnote{In this paper we
refer to the incident and scattered lepton as ``electron''.} of 27.6\,GeV
energy and protons with energy of 820\,GeV. The data set used corresponds
to a total integrated luminosity of 21\,pb$^{-1}$.

The H1 detector consists of a number of sub-detectors~\cite{H1DET1} providing 
complementary and redundant measurements of various aspects of the final
state of high energy
electron-proton collisions.  The detector components which are most important
for this analysis are the backward calorimeter, SpaCal~\cite{SPAPAP1}, together
with the backward drift chamber, BDC~\cite{BDCPAP}, for identifying the
scattered electron, and the Liquid-Argon (LAr) calorimeter~\cite{LAR} for
the measurement of the hadronic final state.  The central tracking system is
used for the determination of the event vertex and to improve the
hadronic energy measurement by the LAr calorimeter~\cite{FSCOMB}.

The SpaCal is a lead/scintillating-fiber calorimeter covering polar
angles\footnote{The $z$ axis of the right-handed coordinate system used by H1
is defined to lie along the direction of the proton beam with the origin 
at the nominal $ep$ interaction vertex.} in the range ${\rm
153^{\circ}} < \theta < {\rm 177.5^{\circ}}$.  Its electromagnetic part
has a depth of 28 radiation lengths and provides an energy resolution of $\sigma_E/E
\approx {\rm 0.07}/\sqrt{E [\mathrm{GeV}]} \oplus {\rm 0.01}$~\cite{SPAPAP2}.
Remaining leakage of electromagnetic showers and energy depositions of hadrons
are measured in the hadronic part of the SpaCal.  The accuracy of the polar
angle measurement of the scattered electron, using the vertex position
and the BDC (${\rm156^{\circ}} < \theta < {\rm 175^{\circ}}$) is 0.5~mrad~\cite{F297}.

The LAr calorimeter covers the angular region ${\rm 4^{\circ}} < \theta <
{\rm 154^{\circ}}$. Its total depth varies between 4.5 and 8 interaction
lengths, depending on the polar angle. It has an energy resolution of
$\sigma_E/E \approx {\rm 0.50}/\sqrt{E [\mathrm{GeV}]} \oplus {\rm 0.02}$ for
charged pions~\cite{LARHADRES}.  The LAr calorimeter surrounds the central
tracking system, which consists of multi-wire proportional chambers and drift
chambers, providing measurements of charged particles with polar angles of
${\rm 15^{\circ}} < \theta < {\rm 165^{\circ}}$.

The SpaCal and LAr calorimeters are surrounded by a superconducting solenoid,
which provides a uniform field of 1.15~T parallel to the beam axis in the
region of the tracking system, allowing track momentum measurements.

The luminosity is determined from the rate of the Bethe-Heitler
process ($ep\rightarrow ep\gamma$).  The luminosity monitor consists of an
electron tagger and a photon detector, both located downstream of the
interaction point in the electron beam direction.

\section{Selection Criteria}

The analysis is based on a sample of DIS events with a clear multi-jet topology
of the hadronic final state.  The events are characterized by an electron
scattered into the backward calorimeter, SpaCal, and at least two jets 
within the acceptance of the LAr calorimeter. They are
triggered by demanding a localized energy deposition in the SpaCal and by
track requirements, which result in a trigger efficiency of~$(97.3 \pm
0.1)$\%~\cite{ROMAN}.

The scattered electron is identified as the cluster of highest energy,
$E_{e} > 9$~GeV, in the electromagnetic part of the SpaCal.  
In order to select well identified electromagnetic 
showers, a cut of 3.5~cm is applied on the energy weighted radius of 
the selected cluster~\cite{ROMAN}. The energy in the hadronic part of the 
SpaCal within a radius of 15~cm of the shower axis is required to be
less than 0.5~GeV. Moreover, an electron candidate must be associated with a track
segment in the BDC.

The inclusive event kinematics are derived from the energy and polar angle measurements of
the electron candidate.  The kinematic range of the analysis is restricted
to the low-$Q^{2}$, low-$x$ region, $5 < Q^2 < 100$~GeV$^2$ and
$10^{-4} < x < 10^{-2}$.  In addition, the inelasticity $y=Q^{2}/xs$
is restricted to $0.1<y<0.7$, where $\sqrt{s}$ is the $ep$ center-of-mass energy. 
In the acceptance region of the SpaCal, the
restriction $y<0.7$ always corresponds to the requirement $E_{e} > 9$~GeV on the
energy of the scattered electron. The requirement $y>0.1$ ensures a large central track
multiplicity and hence the accurate reconstruction of the event
vertex, which is required to lie
 within \mbox{$|z_{\rm vtx}| < {\rm 35}$~cm}. 
The restriction of the $y$-range also reduces the effects of QED bremsstrahlung.

The requirement of $E_{e} > 9$~GeV suppresses background from photoproduction
processes in which the scattered electron escapes through the beam-pipe but an
electron signal is mimicked by a particle from the hadronic
final state. This background is further 
reduced by demanding $35<\sum_{i} (E_{i}-p_{z,i}) < 70$~GeV. Here the sum runs
over the energies and momenta of all final state particles including the
scattered electron.  For fully reconstructed events, energy and momentum
conservation implies that $\sum_{i} (E_{i}-p_{z,i})$ is equal to twice
the energy of the incident electron beam.

Jets are reconstructed in the hadronic center-of-mass system\footnote{
Variables measured in the hadronic cms are marked by a~`$^{\ast}$'.} using the
longitudinally boost invariant $k_{\perp}$-algorithm~\cite{INVKT} and the
$E_{T}$-recombination scheme.  The axis of each reconstructed jet is required
to be within $-{\rm 1} < \eta = -\ln(\tan \frac{\theta}{2}) < {\rm 2.5}$ to
ensure that the jets are well contained within the acceptance of the LAr
calorimeter.  Finally, a minimum transverse jet energy, $E^{\ast}_{T}$, of
5~GeV is required.  Demanding events with at least two jets which fulfill the
criteria listed above yields a total sample of $\sim$~36~000
inclusive dijet events.

\section{Theoretical Predictions}

The available NLO QCD dijet and 3-jet programs provide the partonic final state of the
hard subprocess, to which the chosen jet algorithm and selection can be applied. 
A variety of NLO dijet programs~\cite{DISENT,MIRKES,GRAUDENZ}
have been shown to give comparable results~\cite{CARLO,BJORN,NLOJET}.  Here we use
a slightly modified version of DISENT~\cite{DISENT} in which the renormalization
scale, $\mu^{2}_{r}$, may be set to any linear combination of the two relevant
scales,~$Q^{2}$ and $\bar{{E}^{^\ast}_{T}}^{2}$, where the latter represents the
mean transverse energy squared of the two hardest jets. The
renormalization scale  $\mu^{2}_{r}$ is set to
$\bar{{E}^{^\ast}_{T}}^{2}$ which, for most of the kinematic range
under study, is larger than $Q^{2}$. The factorization scale $\mu^{2}_{f}$ is taken to
be~70~GeV$^2$, {\it i.e.} the average transverse jet energy squared, 
$\langle\bar{{E}^{^\ast}_{T}}^{2}\rangle$, of the event
sample\footnote{DISENT does not allow $\mu^{2}_{f}$ to be varied with
$\bar{{E}^{^\ast}_{T}}^{2}$ event by event.}. The CTEQ6M
(CTEQ6L) PDF parameterizations~\cite{CTEQ6} are used for all NLO (LO)
predictions shown. For NLO 3-jet production, the program NLOJET~\cite{NLOJET} is used.

Theoretical predictions beyond the DGLAP collinear approach, which incorporate
\mbox{low-$x$} effects by assuming different dynamics for the exchanged parton
cascade, are available in Monte Carlo event generators. CCFM evolution,
based on $k_{t}$ factorized unintegrated parton
distributions, is implemented in the CASCADE generator~\cite{CASCADE} for
initial state gluon showers.  An alternative approach is provided by the ARIADNE
Monte Carlo~\cite{ARIADNE} program, which generates non-$k_{t}$-ordered parton
cascades based on the color dipole model~\cite{CDM}.  A LO Monte Carlo
prediction, including effects due to the resolved hadronic structure of the
virtual photon, and generating $k_t$-ordered parton cascades as in the standard
DGLAP approximation, is provided by RAPGAP~\cite{RAPGAP}.  RAPGAP can be run
with (`direct$+$resolved') and without (`direct') a resolved photon
contribution and the data are compared with both scenarios. RAPGAP (`direct') thus
also allows a comparison with the standard DGLAP approach including full
simulation of the hadronic final state.

The LEPTO~\cite{LEPTO} Monte Carlo program, which models only direct photon processes within
the standard DGLAP approximation, and ARIADNE are used to estimate the
hadronization corrections to be applied to the NLO predictions.
All Monte Carlo models used here fragment the partonic final state according to
the LUND string model~\cite{LUND} as implemented in
JETSET/PYTHIA~\cite{JETSET}.

Higher order QED corrections are simulated using HERACLES~\cite{HERACLES}, which
is directly interfaced to RAPGAP and via the DJANGO~\cite{DJANGO} program
to ARIADNE. Both RAPGAP (direct) and ARIADNE are used to estimate the
corrections for QED radiation and for detector effects as is outlined in the next section.

The Monte Carlo programs used in the analysis 
are summarized in Table\,\ref{tab=MC},
including their basic settings. The background contribution from
photoproduction events is estimated with the
PHOJET~\cite{PHOJET} Monte Carlo program.

\begin{table}[t]
\begin{center}
{\footnotesize
\begin{tabular}{l@{\hspace{.4cm}}c@{\hspace{.4cm}}c@{\hspace{.4cm}}c@{\hspace{.4cm}}c}
  &  {\bf CASCADE} & {\bf ARIADNE} &  {\bf RAPGAP}  
  &  {\bf LEPTO} \rule[-2mm]{0mm}{5mm} \\
\hline
\hline
Version       &   1.0    &   4.10   &   2.8   
              &   6.5  \rule[-1mm]{0mm}{5mm} \\
\hline
Proton PDF    &  JS2001~\cite{CASCADE}   
              &  CTEQ5L~\cite{CTEQ5L} &  CTEQ5L        
              &  CTEQ5L \rule[-1mm]{0mm}{5mm}\\
              &  J2003~\cite{JUNGDIS03}    
              &   {}           
              &   {}       
              &   {} \\
Photon PDF    &   {}           &  {}            
              &  SAS1D~\cite{SAS1D} &  {} \rule[-2mm]{0mm}{5mm}\\
\hline
Renorm. scale $\mu_r^2$ \rule[-1mm]{0mm}{6mm} &
                 ${{p}^{^\ast}_{T}}^{2}+m_q^2$ &
                 ${{p}^{^\ast}_{T}}^{2}$ & 
                 ${Q^2+4 {p}^{^\ast}_{T}}^{2}$ &
                 $Q^{2}$ \\
Factor. scale $\mu_f^2$ \rule[-3mm]{0mm}{6mm}&
                 {given by ang. ordering} &
                 ${{p}^{^\ast}_{T}}^{2}$ & 
                 ${Q^2+4 {p}^{^\ast}_{T}}^{2}$ &
                 $Q^{2}$ \\
\hline 
\raisebox{-0.2mm}{Underlying} \rule[1mm]{0mm}{3mm} & 
\raisebox{-0.2mm}{CCFM}        & 
\raisebox{-0.2mm}{Color}       & 
\raisebox{-0.2mm}{DGLAP}      & 
\raisebox{-0.2mm}{DGLAP}    \\
\raisebox{+0.2mm}{model}      \rule[-2mm]{0mm}{3mm} & 
\raisebox{+0.2mm}{}   & 
\raisebox{+0.2mm}{dipole model} & 
\raisebox{+0.2mm}{+ $\gamma$-structure}   & 
\raisebox{+0.2mm}{} \\
\hline         
        & Model comp. 
        & Model comp. 
        & Model comp. & {} \rule[1mm]{0mm}{3mm} \\ 
\raisebox{1.5ex}[-1.5ex]{Purpose}         
        & {}  
        & QED/had. corr. 
        & QED corr.
        & had. corr. \rule[-2mm]{0mm}{3mm} \\
        & {}  
        & detector corr. 
        & detector corr.
        & {} \rule[-2mm]{0mm}{3mm} \\
\hline
\hline
\end{tabular}
\caption{Monte Carlo programs employed in the analysis.}
\label{tab=MC}}
\end{center}
\end{table}

\section{Correction Procedure \label{dcorr}}

In order to compare data with theoretical predictions, the measured cross
sections are corrected for detector acceptance and
resolution, QED radiative effects and background contamination. In addition,
hadronization corrections are applied to the NLO QCD calculations. The various
correction factors are determined using the Monte Carlo models described above.
These models reproduce the gross features of the jet data, as well as
many characteristics of the final state, as shown in~\cite{ROMAN}. However, none
of the models gives a satisfactory description of all aspects of the hadronic
final state. The most important discrepancies are found in the jet
transverse momentum spectra. Differences between the Monte Carlo
models are used to estimate the systematic uncertainties of this procedure.

The corrections are applied to the data after statistical subtraction of
the remaining photoproduction background.  This contamination
is estimated using the PHOJET Monte Carlo program.  It is concentrated in the
low-\myx region and is everywhere less than 4\%.

Detector and QED corrections are estimated using events generated with
ARIADNE and RAPGAP (direct) and subjected to a full detector
simulation and event reconstruction.
The final correction factors are taken to be the average
of the estimates from these two models. Half of the difference 
is included in the systematic uncertainty of the
measurement. For the chosen bins 
purities and stabilities are better than 40\% for all data points. 
Here, the purity (stability) is defined as the number of dijet events
which are both generated and reconstructed in a specific analysis bin, divided
by the total number of dijet events that are reconstructed (generated) in that
bin. The correction factors are in general between 0.8 and 1.2, but reach 1.8 at
the lowest $x$ and $Q^2$ values due to acceptance constraints in the backward
calorimeter~\cite{ROMAN}.  Additional minor corrections are applied to account
for trigger inefficiencies.

As mentioned before, hadronization corrections to the DISENT and NLOJET
predictions are estimated using LEPTO and ARIADNE. The correction factors
are determined by comparing the cross sections calculated from the hadronic
final state (hadron level) with those predicted from the partonic final
state (LO and QCD parton showers) prior to the hadronization step. 
They are obtained by taking the average of the estimates derived from LEPTO and from
ARIADNE. When applied to the NLO predictions, these corrections
allow for comparisons between data and theory at the hadron level. The correction
factors lower the NLO predictions by typically 10\%.  
Half of the difference between the two models is taken as the systematic error on the
hadronization correction.

\section{Systematic Uncertainties \label{sys}}

The different error sources and the corresponding 
uncertainties on the dijet cross section measurements are summarized in
Table~\ref{tab=errors}.  The theoretical uncertainties on the NLO predictions,
given by the errors on the hadronization corrections and the renormalization
scale uncertainty, are also listed.  The latter is 
estimated by varying $\mu^2_r$ between $\bar{{E}^{^\ast}_{T}}^{2}/{\rm 4}$ and ${\rm 4}\,
\bar{{E}^{^\ast}_{T}}^{2}$.  

\begin{table}[t]
\begin{center}
\begin{tabular}{ll@{\hspace{1.cm}}c@{\hspace{.7cm}}c}
 \multicolumn{2}{l}{\bf Source of error contributions} &
 {\bf Variation}  & {\bf Uncertainty} \rule[-2mm]{0mm}{5mm}\\ 
\hline
\hline
Experimental &
  Hadronic energy scale     & 4\%      & 7\% \rule[1mm]{0mm}{3mm}\\
 & SpaCal electromagnetic energy scale  & 1\%      & 5\% \\
 & SpaCal hadronic energy scale  & 7\%      & 2\% \\
 & Electron Polar angle measurement   & 0.5~mrad   & 2\% \\
 & Model uncertainty         &  ---          & 5 -- 10\% \\
 & Photoproduction background    & 30\%     & 1\% \\
 & Normalization uncertainty &  ---          & 1.5\% \rule[-2mm]{0mm}{5mm}\\
\hline
Theoretical &
  Hadronization corrections &  ---          & 5\% \rule[1mm]{0mm}{3mm}\\
&  Renormalization scale uncertainty & ---   & 10-30\% \rule[-2mm]{0mm}{5mm}\\
\hline
\hline
\end{tabular}
\caption{ Summary of error contributions and the resulting typical uncertainties on
the dijet cross section measurements (experimental) and the NLO predictions
(theoretical).}
\label{tab=errors}
\end{center}
\end{table}

One of the most important error contributions arises from the
uncertainty in the hadronic energy measurement used in the jet reconstruction. This scale
uncertainty was estimated to be 4\% and leads to an uncertainty of
typically 7\% on
the dijet cross section measurement, with values increasing up to 20\% at large transverse
jet energies.  The uncertainty of the electromagnetic energy scale of the SpaCal
is 1\% and leads to an error on the dijet cross sections of 5\% in most of
the phase space, reaching $\sim$10\% at large $x$, where in some bins it
constitutes the largest contribution to the total systematic error.  The
influence of the hadronic energy scale uncertainty of the SpaCal of 7\% is of
minor importance. It only enters in the determination of $\sum_{i}
(E_{i}-p_{z,i})$ and gives a 2\% contribution to the final measurement
error.  An error of similar size arises from the polar angle measurement
of the scattered electron.

The differences between the correction factors when using different Monte Carlo
models lead to an error contribution of $\sim$~5 to 10\% throughout the analyzed
phase space.  The 30\% uncertainty on the absolute normalization of the $\gamma
p$-background contributes up to 1\% to the systematic error on the dijet cross
sections.

The total systematic error is determined by summing the individual contributions
in quadrature.  A 1.5\% normalization uncertainty due to the luminosity
measurement is not included in the quoted systematic errors on the
cross sections presented here.

\section{Results \label{results}}

\subsection{Inclusive Dijet Cross Sections}

\begin{figure}[t]
\center
\epsfig{file=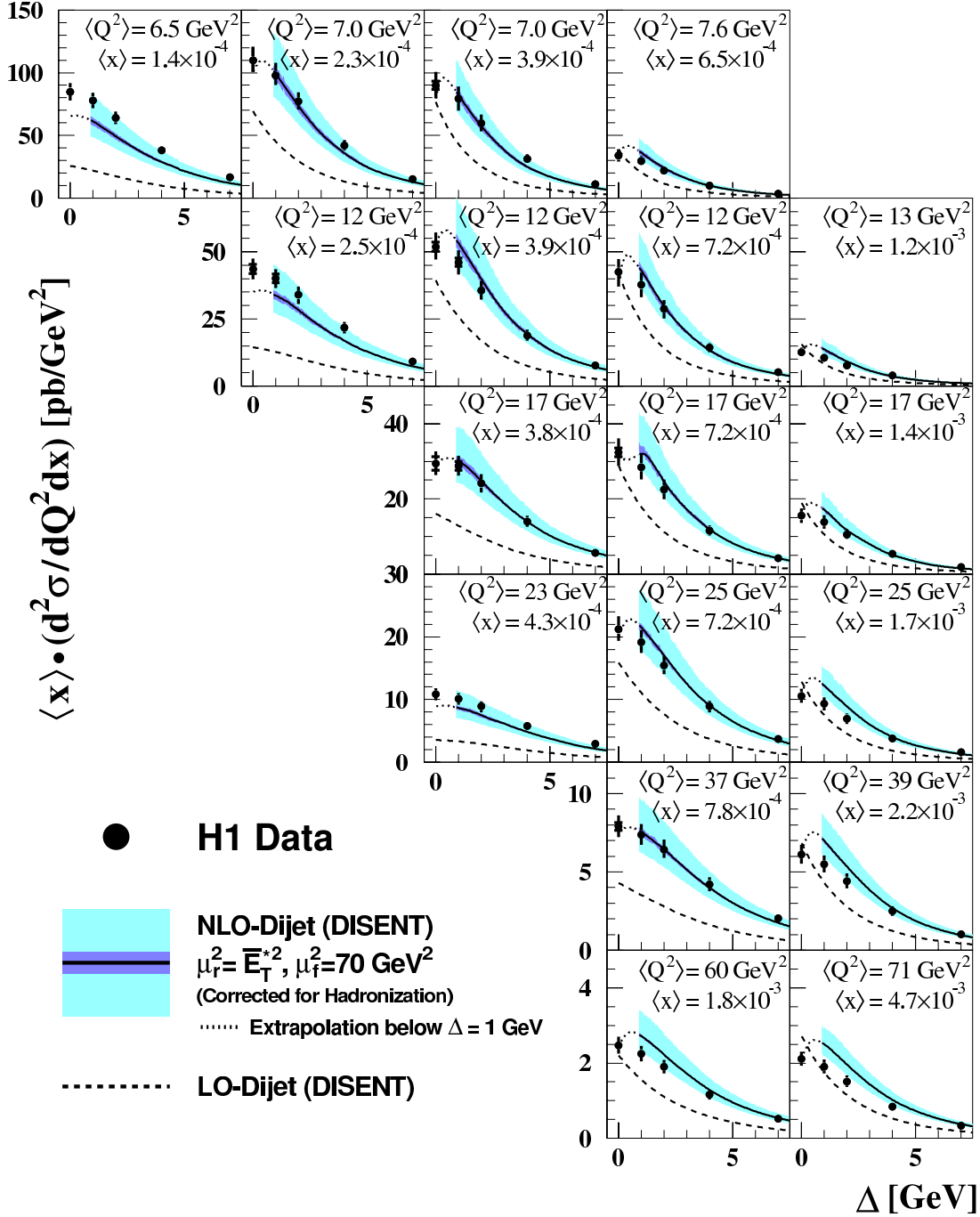,width=0.95\textwidth}
\caption{ 
Inclusive dijet cross section as given in Table~\ref{tab:ddxsec},
multiplied by $\langle x\rangle$ and averaged over \myx and
$Q^{2}$, as a function of $\Delta$, defined by the requirement 
$E^{\ast}_{T,1} \equiv E^{\ast}_{T,{\rm max}} > ({\rm 5}+\Delta)$~GeV. 
Here $\langle x\rangle$ and $\langle Q^2\rangle$ are the mean values of Bjorken-$x$
and $Q^2$ for fully inclusive events in a given bin. The
data are shown together with their statistical uncertainties (inner error bars) and their 
statistical and systematic uncertainties added in quadrature (outer error bars).
They are compared with NLO (LO) dijet QCD predictions using the CTEQ6M (CTEQ6L) 
parton distribution functions. The NLO predictions are corrected for 
hadronization effects. The outer light error band includes the quadratic sum of
hadronization (dark error band) and renormalization scale
uncertainties on the NLO predictions.}
\label{fig:stamp} 
\end{figure}

All measured dijet cross sections are presented after correcting for detector
and radiative effects.  They are given multi-differentially as a function of
$x$, $Q^2$ and several dijet observables and are compared with the DISENT NLO
calculations after applying hadronization corrections. NLO
calculations of dijet observables become sensitive to
soft gluon radiation when symmetric selection criteria on the
transverse jet energies are applied~\cite{BJORN,KRAMKLAS,FRIXIONE}. Thus,
in addition to the requirement $E^{\ast}_{T} > 5$~GeV for the two
highest transverse momentum jets, an additional
requirement on the most energetic jet, $E^{\ast}_{T,1} \equiv E^{\ast}_{T,{\rm
max}} > ({\rm 5}+\Delta$)~GeV, is necessary. This avoids regions of
phase space in which NLO predictions become unreliable.
Figure~\ref{fig:stamp} shows the dijet cross section as a
function of the parameter $\Delta$ in bins of $x$ and $Q^2$. 
Within the theoretical uncertainties, good agreement between the data and the NLO predictions
is found for all values of $\Delta$. However, an unphysical reduction
in the NLO calculation occurs for $\Delta < 1$~GeV.
For comparison the figure also presents the LO DISENT prediction
at the parton level. The large differences between the LO and the NLO predictions, as well
as the large scale uncertainties in the NLO predictions, indicate the need for higher order
contributions, especially at low $x$ and low $Q^2$.

\begin{figure}[t]
\center
\epsfig{file=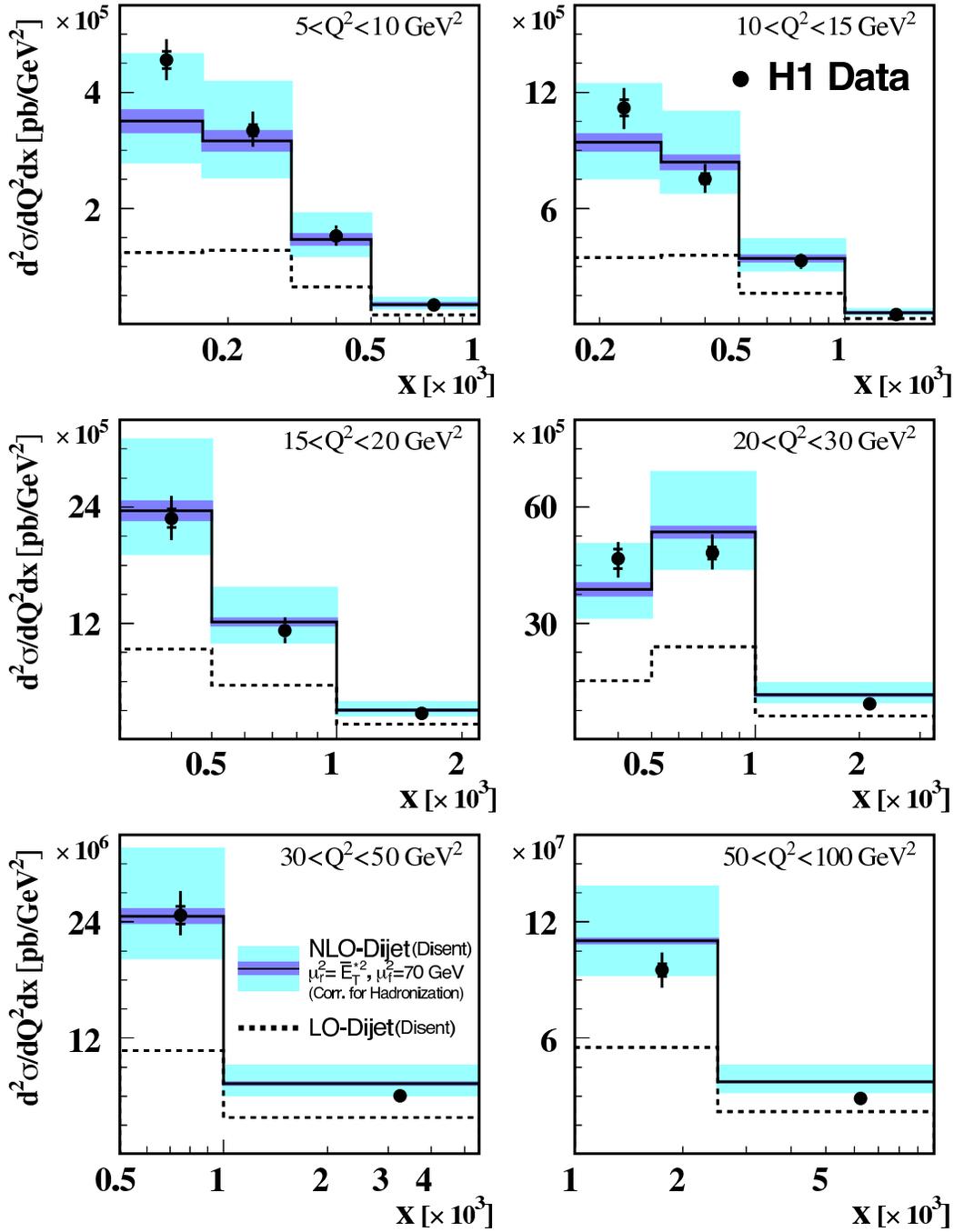,width=0.9\textwidth}
\caption{ 
Inclusive dijet cross section averaged over \qsq and Bjorken-$x$ for
$\Delta=2$~GeV (see text). The data are plotted at the center of each
bin and are shown with their
statistical uncertainties (inner error bars) and their statistical and systematic 
uncertainties added in quadrature (outer error bars). They are compared with NLO
(LO) dijet QCD predictions using the CTEQ6M (CTEQ6L) parton distribution
functions. The NLO predictions are corrected for 
hadronization effects. The outer light error band includes the quadratic sum of
hadronization (dark error band) and renormalization scale
uncertainties on the NLO predictions. 
}
\label{fig:dtmodnlo} 
\end{figure}

Figure~\ref{fig:dtmodnlo} shows the dijet cross section as a
function of Bjorken-$x$ in intervals of $Q^2$ for fixed $\Delta =
$~2~GeV. 
The data show a significant increase towards low $x$, which is consistent
with the strong rise of the gluon density observed in low-$x$ structure function
measurements at HERA~\cite{F297, Chekanov:2002pv}.  No deviation of
the data from calculations using  the conventional NLO DGLAP approach is
found.\footnote{Minor
differences at low $x$ and $Q^2$ as reported in~\cite{r2js} are
still observed when using CTEQ4M~\cite{CTEQ4M}, an older
parameterization of the parton distribution functions.} The scale
uncertainties are sizable and increase towards low $x$.  

\begin{figure}[t]
\center
\epsfig{file=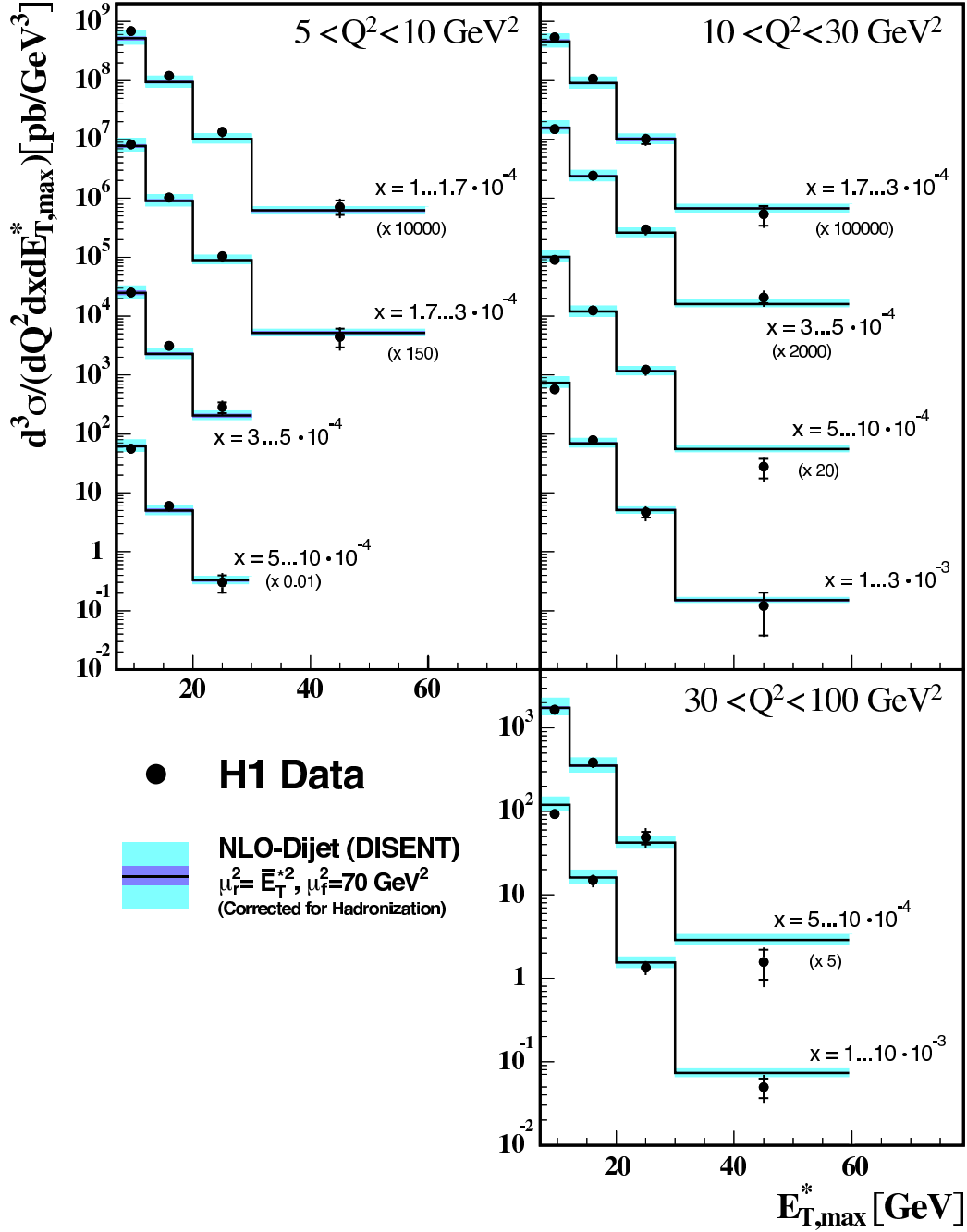,width=0.95\textwidth}
\caption{
Inclusive dijet cross section for $\Delta = 2$~GeV averaged over Bjorken-$x$,
$Q^2$ and $E^{\ast}_{T,{\rm max}}$ as given in Table~\ref{tab:xsecpt},
compared with NLO dijet 
QCD predictions using the CTEQ6M parton distribution functions. The data are
plotted at the center of each bin and are shown together with their statistical 
uncertainties (inner error bars) and their statistical and systematic uncertainties added in 
quadrature (outer error bars).  
The NLO predictions are corrected for 
hadronization effects. The outer light error band includes the quadratic sum of
hadronization (dark error band) and renormalization scale
uncertainties on the NLO predictions. 
}
\label{fig:ptjet} 
\end{figure}

\begin{figure}[t]
\center
\epsfig{file=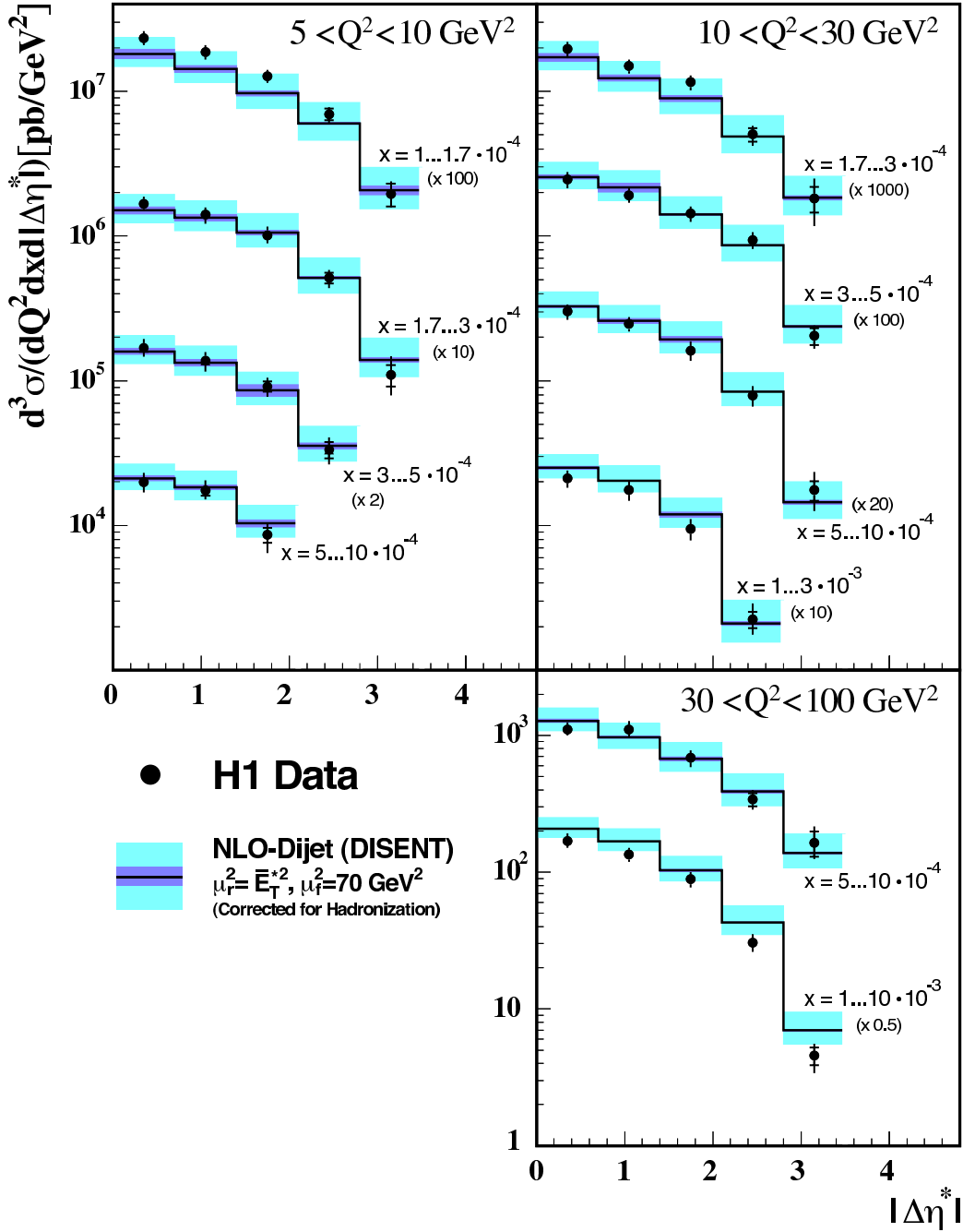,width=0.95\textwidth}
\caption{
Inclusive dijet cross section for $\Delta = 2$~GeV averaged over Bjorken-$x$,
$Q^2$ and the pseudorapidity distance $|\Delta\eta^{\ast}|$ between the dijets 
as given in Table~\ref{tab:xseceta}, compared with NLO dijet QCD
predictions using the CTEQ6M parton distribution 
functions. The data are plotted at the center of each bin and are shown together
with their statistical uncertainties (inner error bars) and their statistical and systematic
uncertainties added in quadrature (outer error bars).  The NLO predictions are 
corrected for hadronization effects. The outer light error band includes the quadratic 
sum of hadronization (dark error band) and renormalization scale
uncertainties on the NLO predictions.}
\label{fig:etajet} 
\end{figure}

\begin{figure}[t]
\center
\epsfig{file=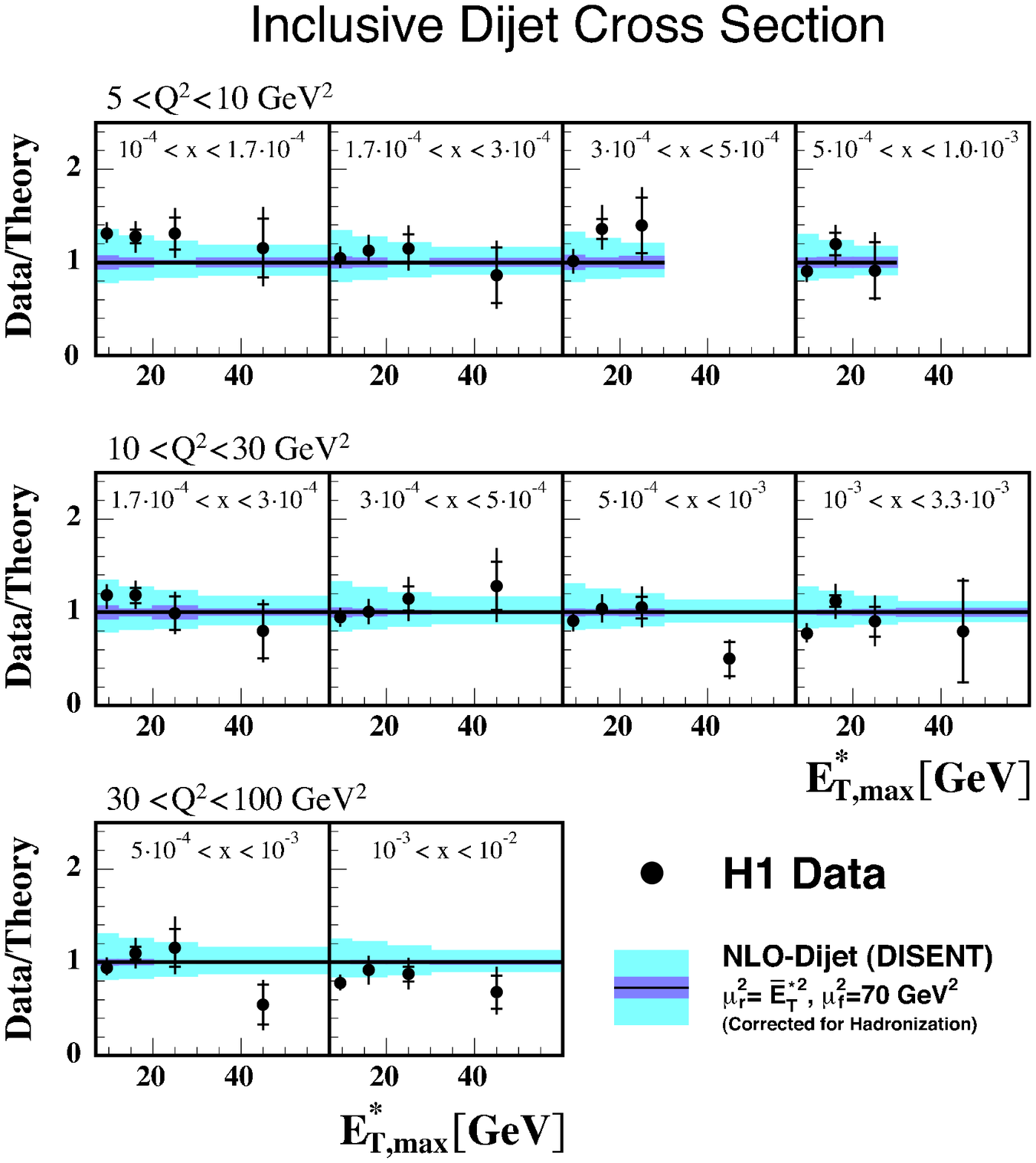, width=0.95\textwidth}
\caption{
The ratio of the measured inclusive dijet cross section for $\Delta =
2$~GeV to the
theoretical prediction in bins of Bjorken-$x$,
\qsq and $E^{\ast}_{T,{\rm max}}$. The data are shown together with their statistical
uncertainties (inner error bars) and their statistical and systematic uncertainties added in
quadrature (outer error bars). They are compared with NLO dijet QCD
predictions using the CTEQ6M parton distribution functions. The
theoretical errors are given by the light error band 
representing the quadratic sum of the hadronization (dark error band) and 
renormalization scale uncertainties.}
\label{fig:ptcomp} 
\end{figure}

\begin{figure}[t]
\center
\epsfig{file=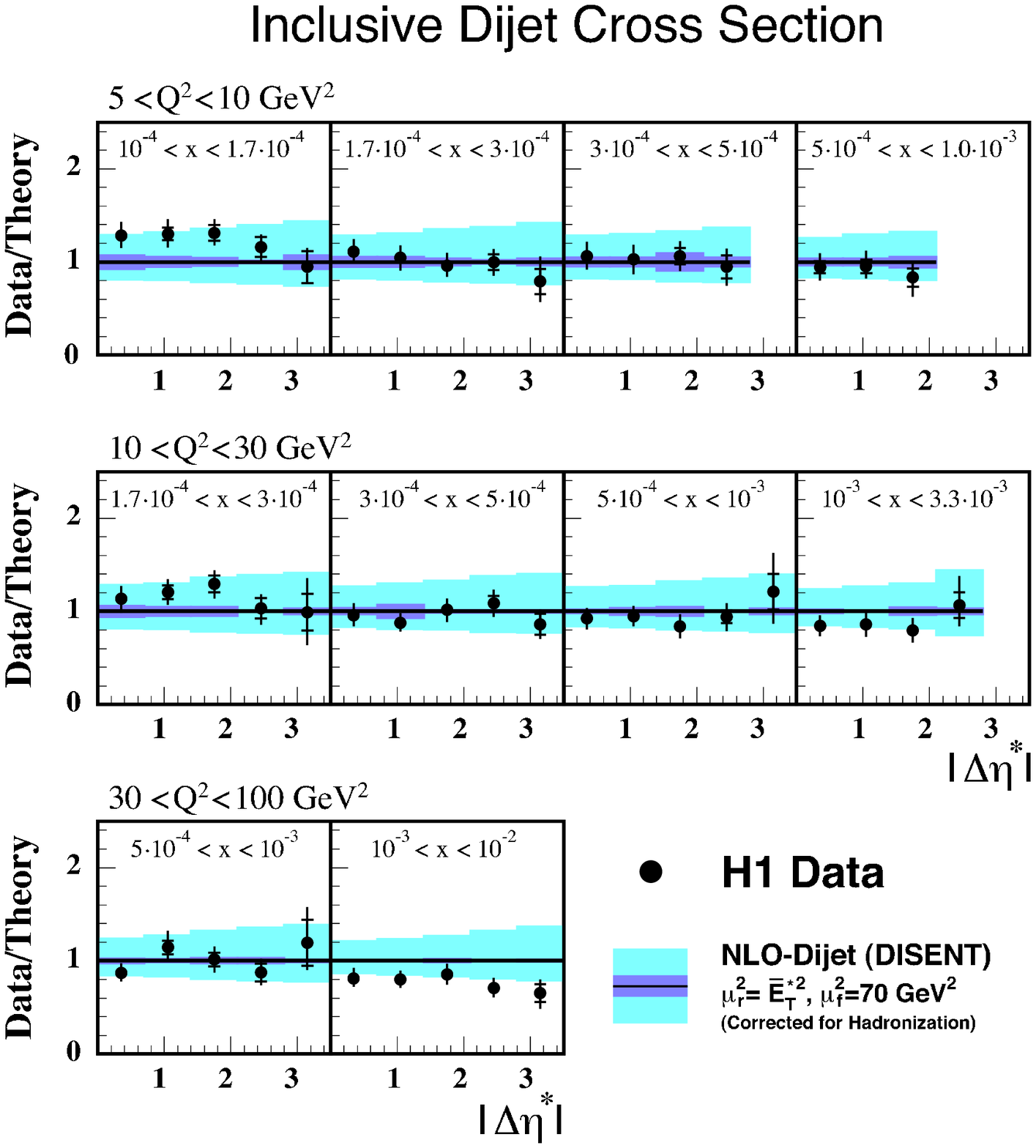,width=0.95\textwidth}
\caption{
The ratio of the measured inclusive dijet cross section for $\Delta = 2$~GeV to the
theoretical prediction in bins of Bjorken-$x$, 
\qsq and $|\Delta\eta^{\ast}|$. The data are shown together with their statistical
uncertainties (inner error bars) and their statistical and systematic uncertainties added in
quadrature (outer error bars). They are compared with NLO dijet QCD 
predictions using the CTEQ6M parton distribution functions. The theoretical errors are given by the light error band 
representing the quadratic sum of the hadronization (dark error band) and 
renormalization scale uncertainties. }
\label{fig:etacomp} 
\end{figure}

Measurements of the dijet cross section for $\Delta = 2$~GeV as a function of
$E^{\ast}_{T,{\rm max}}$ and $|\Delta\eta^{\ast}|$ in bins of
Bjorken-$x$ and $Q^2$ are shown in Figures~\ref{fig:ptjet} and~\ref{fig:etajet}. 
Within the quoted uncertainties, good agreement between data and
NLO calculations is observed, even at small values of $E^{\ast}_{T,{\rm
max}}$ and $|\Delta\eta^{\ast}|$, {\it i.e.} in a kinematic region in which effects
due to low-$x$ dynamics should be most prominent~\cite{ASKE}.
The level of agreement is more visible in Figures~\ref{fig:ptcomp}
and~\ref{fig:etacomp} which show the ratio of the data to the NLO predictions as a
function of $E^{\ast}_{T,{\rm max}}$ and $|\Delta\eta^{\ast}|$ in bins
of Bjorken-$x$ and $Q^2$.

The measured dijet cross sections are summarized in 
Tables\,\ref{tab:ddxsec} to \ref{tab:xseceta}.  In
addition, all dijet cross sections shown have been normalized to the total
inclusive cross section for each bin. These dijet rates, $R_2 = N_{\rm
dijet}/N_{\rm DIS}$, are listed in Tables\,\ref{tab:r2sa} to \ref{tab:r2eta}.

\subsection{Azimuthal Jet Separation}

\begin{figure}[ht]
\center
\epsfig{file=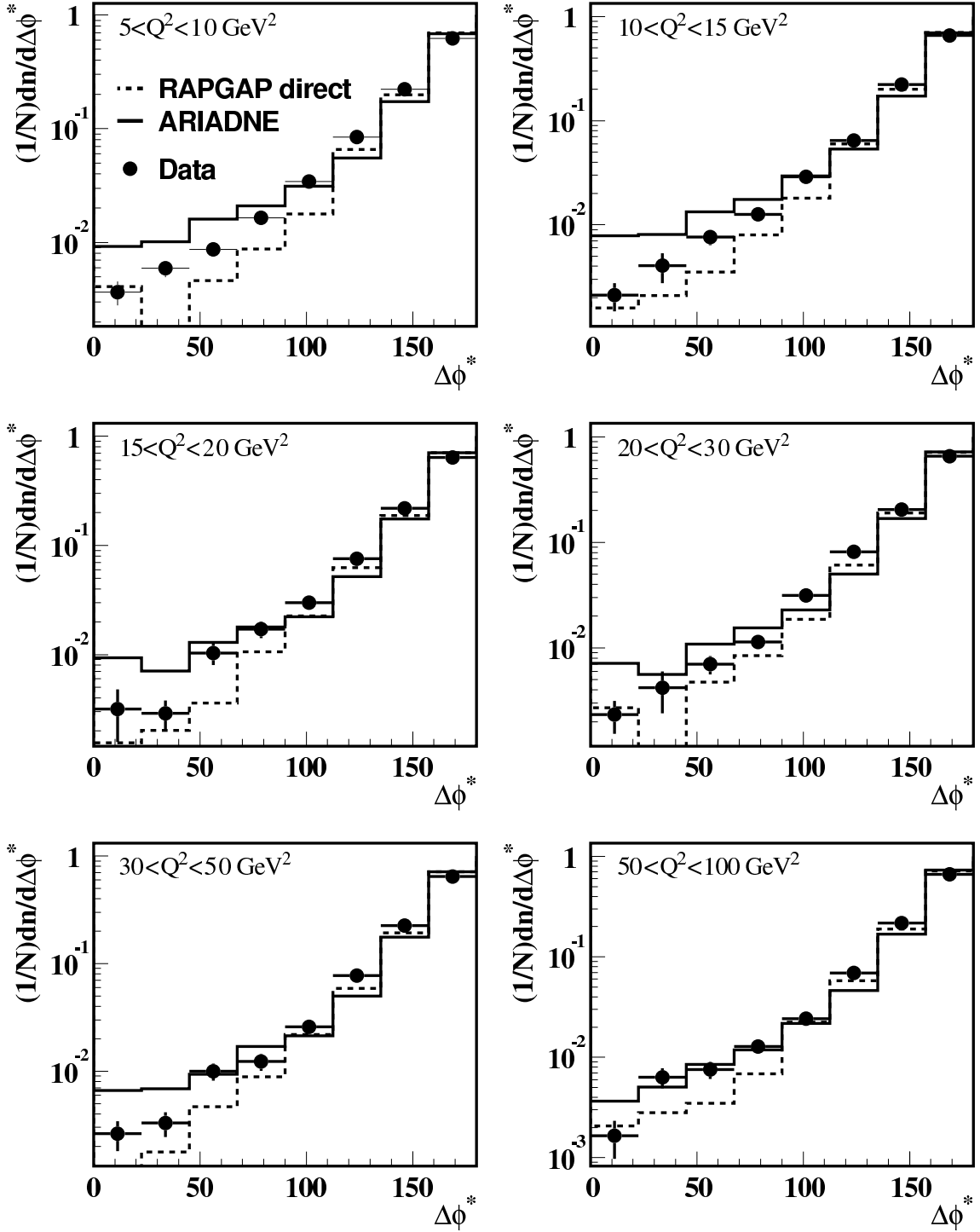,width=0.95\textwidth}
\caption{\label{fig:phiplot}
Normalized measured event distributions as a function of the 
azimuthal separation, $\Delta\phi^*$, between the two highest transverse momentum jets.
The data are shown for different regions in $Q^2$ and are
compared with two Monte Carlo models, ARIADNE and RAPGAP.} 
\end{figure}

Insight into low-$x$ dynamics can be gained from inclusive dijet data
by studying the behavior of events with a small azimuthal separation,
$\Delta\phi^{\ast}$, between the two hardest jets as measured in the hadronic
center-of-mass system~\cite{FORS, ASKE, SZCU}.  Partons entering the hard
scattering process with negligible transverse momentum, $k_t$, as assumed in the
DGLAP formalism, produce at leading order a back-to-back configuration of the two outgoing
jets with $\Delta\phi^{\ast} \sim 180^{\circ}$.  Azimuthal jet separations different from
$180^{\circ}$ occur due to higher order QCD effects.  However, in models which
predict a significant proportion of partons entering the hard process with large
$k_t$, the number of events with small $\Delta\phi^{\ast}$ increases.  This
is the case for the BFKL and CCFM evolution schemes.  As an illustration,
Figure~\ref{fig:phiplot} shows the uncorrected $\Delta\phi^{\ast}$ distribution
for several intervals in $Q^2$. The expected steeply falling
spectrum is observed with a tail extending to small values of
$\Delta\phi^{\ast}$. This behavior is broadly reproduced by the Monte
Carlo programs RAPGAP (direct) and ARIADNE, which are used to correct
the data and estimate model uncertainties in the same manner as 
for the differential cross section measurements.

Large migrations connected with the limited hadronic energy resolution
make an extraction of the dijet cross section at small $\Delta\phi^{\ast}$ rather difficult. 
Thus the ratio
\begin{equation*}
S(\alpha)=\frac{\int^{\alpha}_{0}{N_{\rm dijet}(\Delta\phi^{*}, \myx,
\qsq){\rm d}\Delta\phi^{*}}} {\int^{180^{\circ}}_{0}{N_{\rm dijet}(\Delta\phi^{*}, \myx,
\qsq){\rm d}\Delta\phi^{*}}}, 
\label{sasi}
\end{equation*}
of the number of events $N_{\rm dijet}$ with an azimuthal jet separation of
$\Delta\phi^{\ast}< \alpha$ relative to all dijet events
is measured, as proposed in~\cite{SZCU}. 
This variable is also directly sensitive to low-$x$ effects.  For the
analysis presented here $\alpha = 120^{\circ}$ is chosen, which results in a
purity of around 45\% independently of $x$ and $Q^2$ and systematic
uncertainties of similar size to those on the
cross section measurements.

\begin{figure}[t]
\center
\epsfig{file=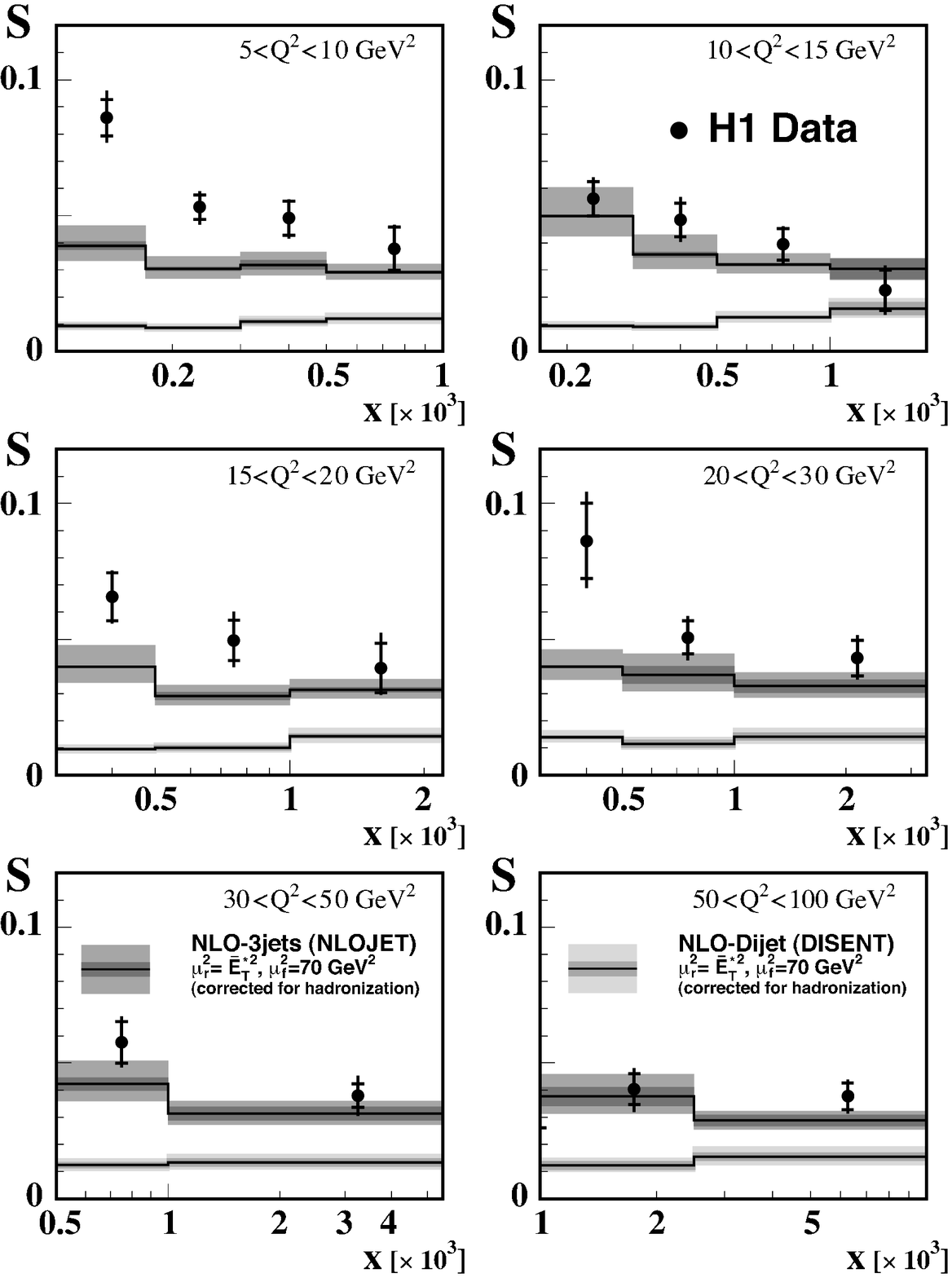,width=0.85\textwidth} 
\caption{
Ratio $S$ of the number of events with a small azimuthal jet separation
($\Delta\phi^{\ast}<120^{\circ}$) between the two highest transverse
momentum jets with respect to the total
number of inclusive dijet events, as a function of Bjorken-$x$ and $Q^2$.
The data are plotted at the center of each bin and are 
shown together with their statistical uncertainties (inner error bars) and their
statistical and systematic uncertainties added in quadrature (outer error bars).
The data are compared with NLO QCD predictions for dijet and 3-jet production
using the CTEQ6M parton distribution functions. The
theoretical errors are given by the light error band representing the quadratic
sum of the hadronization (dark error band) and renormalization scale 
uncertainties. }
\label{fig:ssjet120} 
\end{figure}

Figure~\ref{fig:ssjet120} presents the $S$ distribution for
$\alpha=120^{\circ}$ as a function
of $x$ for different $Q^2$. The measured $S$ values are summarized in
Table~\ref{tab:ssjets}. For the chosen $\alpha$, the measured
values of $S$ are of the order of~5\% and increase with decreasing $x$.  This rise of $S$
is most prominent in the lowest \qsq bin, where the lowest values of \myx are reached. 
The NLO dijet QCD calculations predict $S$ values of only~$\sim$1\% and show no
rise towards low $x$.  Low values of $S$ are expected for the NLO
dijet predictions, since without any restrictions in acceptance, 
the two most energetic jets should always be separated by more than
$\Delta\phi^{\ast}=120^{\circ}$. However, since 
selection criteria have to be applied to match the experimental conditions,
non-vanishing $S$ values arise, due to event topologies for which some of the
jets lie outside the analyzed phase space.  
In the same figure NLO 3-jet predictions
are also shown. These give a good description of the data at large $Q^2$ and large $x$,
but still fail to describe the increase towards low $x$, particularly in the
lowest $Q^2$ range.

\begin{figure}[t]
\center \epsfig{file=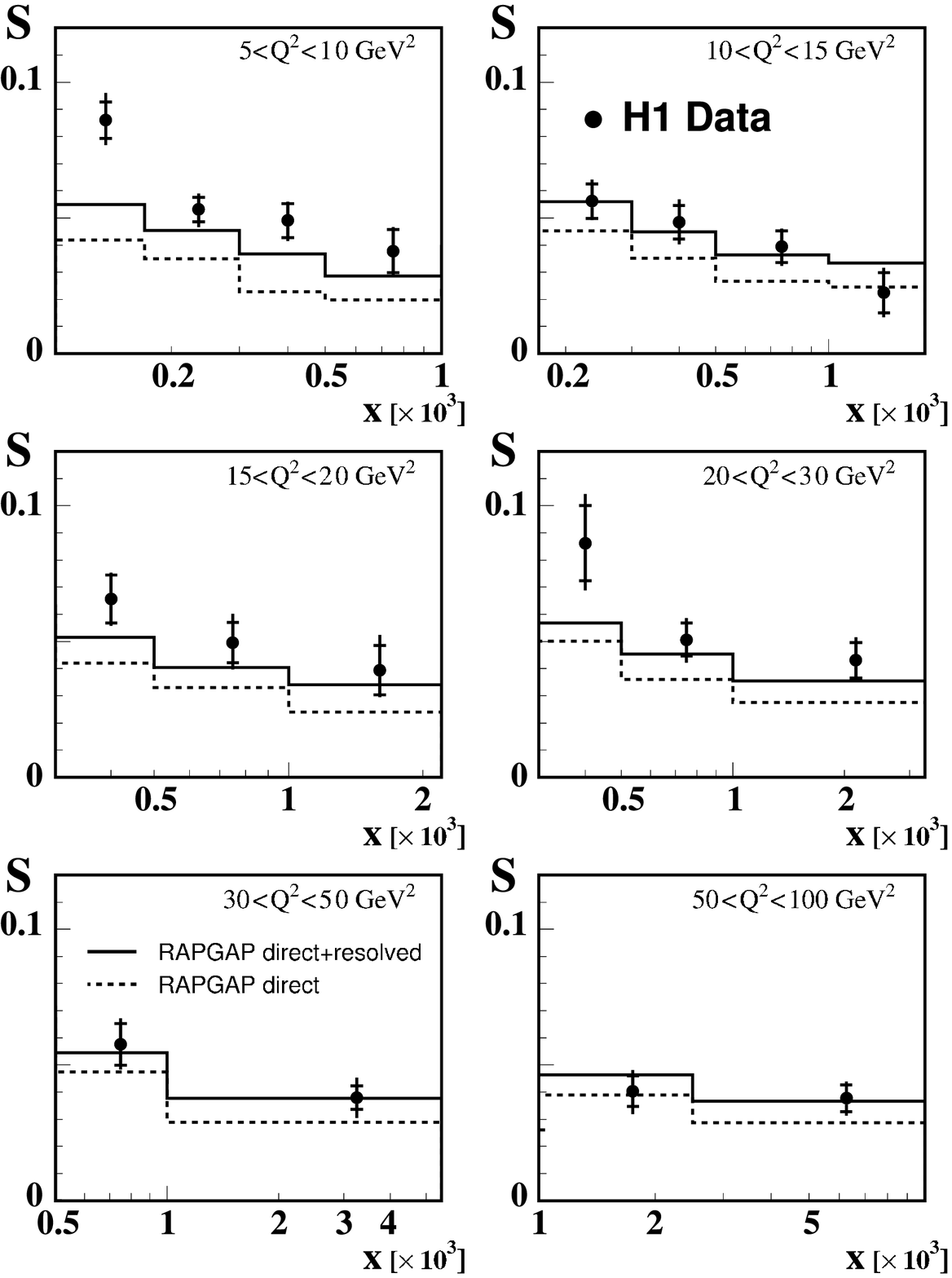,width=0.85\textwidth}
\caption{ 
Ratio $S$ of the number of events with a small azimuthal jet separation
($\Delta\phi^{\ast}<120^{\circ}$) between the two highest transverse
momentum jets with respect to the total
number of inclusive dijet events, as a function of Bjorken-$x$ and $Q^2$.
The data are plotted at the center of each bin and are shown together with their 
statistical uncertainties (inner error bars) and their statistical and systematic uncertainties
added in quadrature (outer error bars). 
The data are compared with predictions from the RAPGAP generator,
both with direct photons alone (full line) and with direct and resolved contributions
(dashed line). 
}
\label{fig:ssjet120rg} 
\end{figure}

An informative comparison of the measured $S$ distribution with theory is 
also provided by models with different implementations of higher order QCD
effects.
Within the DGLAP approach such a model is provided for example 
by RAPGAP (direct).  
As shown in Figure~\ref{fig:ssjet120rg}, RAPGAP (direct) predicts a
much larger ratio $S$ than the NLO dijet calculation. However, it
still fails to
describe the data in the low-$x$, low-$Q^2$ region. An improved
description is achieved when resolved photon processes are included in RAPGAP
(direct$+$resolved). Even with direct and resolved photon 
contributions included, RAPGAP fails to describe the data at very low $x$ and 
$Q^2$.  Note that to obtain this overall level of agreement, it was necessary to choose a
rather large scale, {\it i.e.} $\mu_r^2 = Q^2 + {\rm 4} \bar{{E}^{^\ast}_{T}}^2$, in 
order to get a large enough resolved photon contribution.

\begin{figure}[t]
\center \epsfig{file=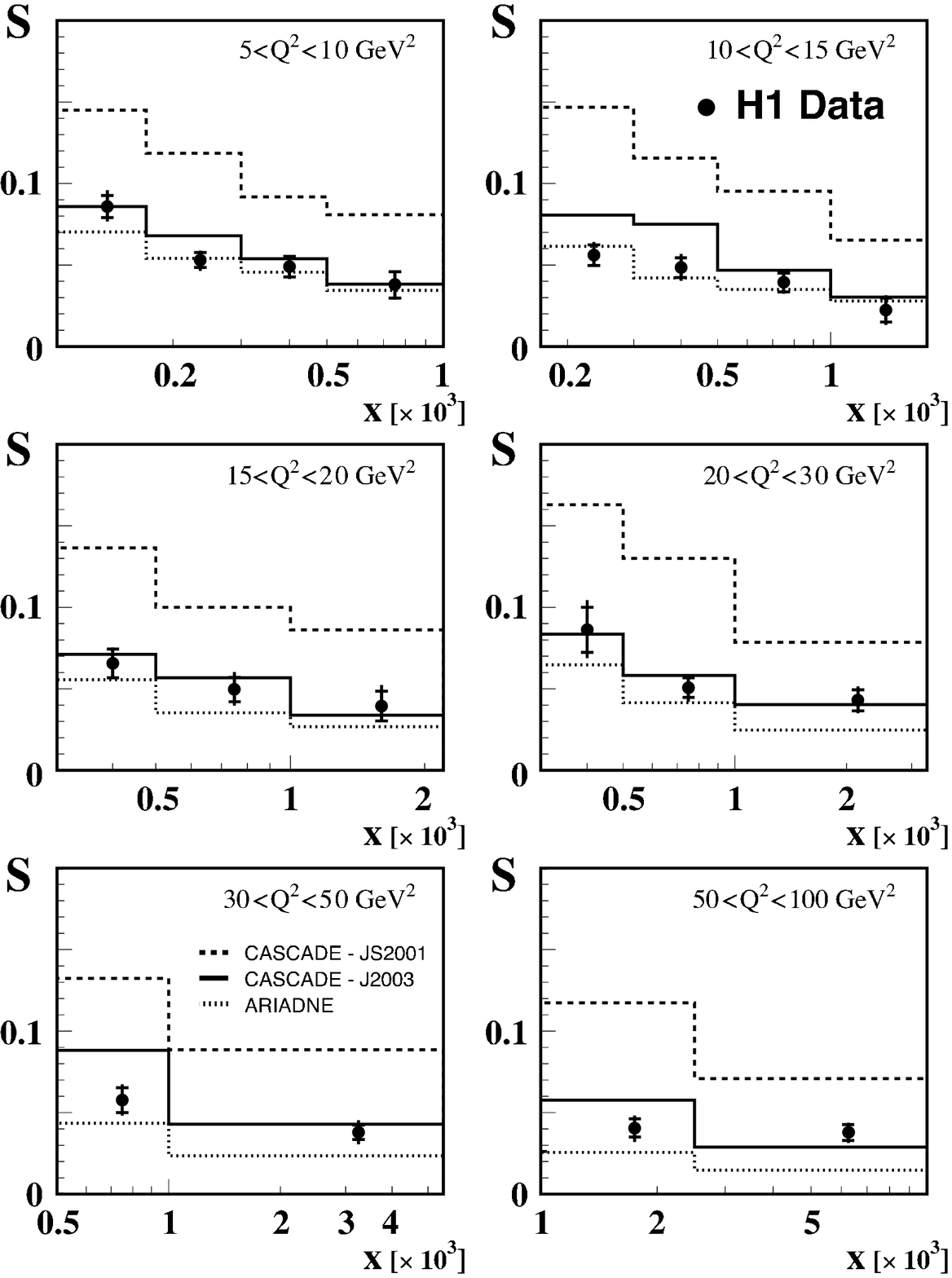,width=0.85\textwidth}
\center 
\caption{ 
Ratio $S$ of the number of events with a small azimuthal jet
separation 
\mbox{($\Delta\phi^{\ast}<120^{\circ}$)} between the two highest
transverse momentum jets with respect to the total
number of inclusive dijet events, as function of Bjorken-$x$ and \qsq.
The data are plotted at the center of each bin and are 
shown together with their statistical uncertainties (inner error bars) and their
statistical and systematic uncertainties added in quadrature (outer error bars).
They are compared with predictions from a model based on CCFM 
evolution (CASCADE) and using two different $k_{t}$-unintegrated gluon
distribution functions, JS2001 (full line) and set 2 of J2003 (dashed line).
In addition, the data are compared with predictions based on the color dipole 
model (ARIADNE), which produces $k_{t}$-unordered parton showers. }
\label{fig:ssjet120casc} 
\end{figure}

If the observed discrepancies are due to the influence of 
non-$k_t$-ordered parton emissions, models based on the color dipole model or
CCFM evolution may provide a better description of the ratio $S$.  In
Figure~\ref{fig:ssjet120casc} the data are therefore compared with the
predictions of the ARIADNE and CASCADE Monte Carlo programs.  For CASCADE, the two
predictions presented are based on the JS2001~\cite{CASCADE} and 
set 2 of the J2003~\cite{JUNGDIS03} unintegrated parton distributions, which
differ in the way the small $k_{t}$ region is treated.  For J2003 set 2,
the full splitting function, including the non-singular term, is used,
in contrast to JS2001, for which only the singular terms were considered. 
Whereas the prediction for $S$ using JS2001 lies significantly above the data,
that based on set 2 of J2003 describes the data rather well.  
Note that both PDFs describe the H1 structure function data~\cite{FSCOMB}.  
Hence, the measurement of the ratio $S$ improves the sensitivity 
to the details of the unintegrated gluon distribution. 
A good description of the $S$ distribution at low $x$ and
low $Q^2$ is also provided by the color dipole model incorporated in ARIADNE.
However, at higher $Q^2$ the ARIADNE prediction falls below the
measured $S$ values.

\section{Conclusion}

Inclusive dijet production in deep inelastic $ep$ scattering is
measured in the kinematic range $5 < Q^2 < 100$~GeV$^2$,  $10^{-4} < x <
10^{-2}$ and $0.1 < y < 0.7$. Multi-differential cross section data
are compared  with NLO QCD predictions and no
significant deviations are observed within the experimental and theoretical
uncertainties.  In the kinematic range studied, the next-to-leading order DGLAP
approach thus provides an adequate theory for predicting $ep$ dijet cross
sections as a function of Bjorken-$x$, $Q^2$, $E^{\ast}_{T,{\rm max}}$ and
$|\Delta\eta^{\ast}|$. 

NLO dijet QCD calculations predict values that are much too low for
the ratio, $S$, of events with a small azimuthal 
separation of the two highest transverse momentum jets with respect to the total number of
inclusive dijet events. The additional hard emission, provided
by the NLO 3-jet calculations, considerably improves the description of the data,
but is insufficient at low $x$ and low $Q^2$.  A similar description
of the data is provided by RAPGAP, a DGLAP-based QCD model,
which matches LO matrix elements for direct and resolved processes
to $k_{t}$-ordered parton cascades.  A good description of the
measured ratio $S$ at low $x$ and $Q^2$ is given by the ARIADNE
program, which generates non-$k_{t}$-ordered parton
cascades using the color dipole model.  
Predictions based on the CCFM evolution equations and $k_t$ factorized
unintegrated gluon densities are provided by the CASCADE Monte Carlo
program. Large differences are found between the predictions for two
different choices of the unintegrated gluon
density, both of which describe the H1 structure function data and one of which
gives a good description of $S$.  This measurement thus provides
a significant constraint on the unintegrated gluon density.

\section*{Acknowledgments}

We are grateful to the HERA machine group whose outstanding efforts have made
this experiment possible.  We thank the engineers and technicians for their work
in constructing and now maintaining the H1 detector, our funding agencies for
financial support, the DESY technical staff for continual assistance and the
DESY directorate for support and for the hospitality which they extend to the
non DESY members of the collaboration.

\clearpage

\begin{center}

\include{ddstablept}

\include{xstablept}

\include{xstableeta}

\include{r2dtablept}

\include{rstablept}

\include{rstableeta}

\include{xstabless}

\end{center}

\end{document}

%% file: h1auts.tex

A.~Aktas$^{10}$,               
V.~Andreev$^{24}$,             
T.~Anthonis$^{4}$,             
A.~Asmone$^{31}$,              
A.~Babaev$^{23}$,              
S.~Backovic$^{35}$,            
J.~B\"ahr$^{35}$,              
P.~Baranov$^{24}$,             
E.~Barrelet$^{28}$,            
W.~Bartel$^{10}$
% ...
\newline
$ ^{\dagger}$ deceased

\bigskip{\it
 $ ^{1}$ I. Physikalisches Institut der RWTH, Aachen, Germany$^{ a}$ \\
 $ ^{2}$ III. Physikalisches Institut der RWTH, Aachen, Germany$^{ a}$ \\
 $ ^{3}$ School of Physics and Space Research, University of Birmingham,
          Birmingham, UK$^{ b}$ \\
 $ ^{4}$ Inter-University Institute for High Energies ULB-VUB, Brussels;
          Universiteit Antwerpen (UIA), Antwerpen; Belgium$^{ c}$ \\
% ...
$ ^{10}$ DESY, Hamburg, Germany \\
% ...
$ ^{23}$ Institute for Theoretical and Experimental Physics,
          Moscow, Russia$^{ l}$ \\
 $ ^{24}$ Lebedev Physical Institute, Moscow, Russia$^{ e}$ \\
% ...
$ ^{28}$ LPNHE, Universit\'{e}s Paris VI and VII, IN2P3-CNRS,
          Paris, France \\
% ...
$ ^{31}$ Dipartimento di Fisica Universit\`a di Roma Tre
          and INFN Roma~3, Roma, Italy \\
% ...
$ ^{35}$ DESY, Zeuthen, Germany \\
% ...
$ ^c$ Supported by FNRS-FWO-Vlaanderen, IISN-IIKW and IWT \\
% ...
$ ^e$ Supported by the Deutsche Forschungsgemeinschaft \\
% ...
$ ^l$ Partially Supported by Russian Foundation
      for Basic Research, grant    no. 00-15-96584 \\
}

%% file: ddstablept.tex
 \begin{table}[htp]
 \footnotesize
 \centering
 \begin{tabular}{|d{-1}cd{-1}|d{-1}cd{-1}|d{-1}|c|cc|}
 \hline
 \multicolumn{3}{|c|}
{\rule[-0.7mm]{0mm}{6.0mm} \footnotesize $Q^{2}$} & 
 \multicolumn{3}{c|}
{\footnotesize $x/10^{-4}$} & 
 \multicolumn{1}{c|}
{\footnotesize $\Delta$} & 
 \multicolumn{1}{c|}
{\normalsize $\frac{d^{2}\sigma}{dQ^{2}dx}$} & 
 \multicolumn{1}{c}
 {\footnotesize $\delta_{stat}$} & 
 \multicolumn{1}{c|}
 {\footnotesize $\delta_{syst.}$} \\
 \multicolumn{3}{|c|}
 {$\rm\ [GeV^{2}]$} &
 \multicolumn{3}{c|} {} &
 \multicolumn{1}{c|} {$\rm\ [GeV]$} &
 \multicolumn{1}{c|} {$\rm\ [pb/GeV^{2}]$} &
 \multicolumn{1}{c} {$[\%]$} & 
 \multicolumn{1}{c|} {$[\%]$} 
 \rule[-2mm]{0mm}{6mm}\\
 \hline
 \rule[-0.1mm]{0mm}{3mm} 5.0&--& 10.0&  1.0&--& 1.7&0&$6.1\cdot10^{5}$&  3 &  8 \\
&&&&&&1&$5.6\cdot10^{ 5}$&  3 &  7 \\
&&&&&&2&$4.6\cdot10^{ 5}$&  3 &  7 \\
&&&&&&4&$2.7\cdot10^{ 5}$&  4 &  8 \\
&&&&&&7&$1.2\cdot10^{ 5}$&  6 &  9\rule[-1.5mm]{0mm}{3.mm}\\
 \hline
&&&  \rule[-0.1mm]{0mm}{3mm} 1.7&--&  3.0&0&$4.8\cdot10^{ 5}$&  3 &  9 \\
&&&&&&1&$4.3\cdot10^{ 5}$&  3 &  10 \\
&&&&&&2&$3.4\cdot10^{ 5}$&  3 &   9 \\
&&&&&&4&$1.8\cdot10^{ 5}$&  4 &   9 \\
&&&&&&7&$6.6\cdot10^{ 4}$&  5 &  11\rule[-1.5mm]{0mm}{3.mm}\\
 \hline
&&&  \rule[-0.1mm]{0mm}{3mm} 3.0&--&  5.0&0&$2.3\cdot10^{ 5}$&  4 & 12 \\
&&&&&&1&$2.0\cdot10^{ 5}$&  4 & 12 \\
&&&&&&2&$1.5\cdot10^{ 5}$&  4 & 11 \\
&&&&&&4&$8.0\cdot10^{ 4}$&  5 & 11 \\
&&&&&&7&$2.9\cdot10^{ 4}$&  8 & 13\rule[-1.5mm]{0mm}{3.mm}\\
 \hline
&&& \rule[-0.1mm]{0mm}{3mm} 5.0&--& 10.0&0&$5.2\cdot10^{ 4}$&  5 & 14 \\
&&&&&&1&$4.6\cdot10^{ 4}$&  5 & 13 \\
&&&&&&2&$3.4\cdot10^{ 4}$&  5 & 13 \\
&&&&&&4&$1.6\cdot10^{ 4}$&  7 & 12 \\
&&&&&&7&$5.4\cdot10^{ 3}$& 10 & 15\rule[-1.5mm]{0mm}{3.mm}\\
 \hline
\rule[-0.1mm]{0mm}{3mm} 10.0&--& 15.0&  1.7&--&  3.0&0&$1.7\cdot10^{5}$&  4 &  8 \\
&&&&&&1&$1.6\cdot10^{ 5}$&  4 &  8 \\
&&&&&&2&$1.4\cdot10^{ 5}$&  4 &  9 \\
&&&&&&4&$8.7\cdot10^{ 4}$&  5 &  9 \\
&&&&&&7&$3.7\cdot10^{ 4}$&  7 &  8\rule[-1.5mm]{0mm}{3.mm}\\
 \hline
&&& \rule[-0.1mm]{0mm}{3mm} 3.0&--&  5.0&0&$1.3\cdot10^{ 5}$&  3 &  10 \\
&&&&&&1&$1.2\cdot10^{ 5}$&  3 &  9 \\
&&&&&&2&$9.1\cdot10^{ 4}$&  4 &  9 \\
&&&&&&4&$4.8\cdot10^{ 4}$&  5 & 10 \\
&&&&&&7&$2.0\cdot10^{ 4}$&  7 & 11\rule[-1.5mm]{0mm}{3.mm}\\
 \hline
&&& \rule[-0.1mm]{0mm}{3mm} 5.0&--& 10.0&0&$5.9\cdot10^{ 4}$&  3 & 13 \\
&&&&&&1&$5.2\cdot10^{ 4}$&  3 & 12 \\
&&&&&&2&$4.0\cdot10^{ 4}$&  4 & 12 \\
&&&&&&4&$2.0\cdot10^{ 4}$&  5 & 12 \\
&&&&&&7&$7.3\cdot10^{ 3}$&  7 & 13\rule[-1.5mm]{0mm}{3.mm}\\
 \hline
&&& \rule[-0.1mm]{0mm}{3mm} 10.0&--& 18.0&0&$1.1\cdot10^{ 4}$&  6 & 12 \\
&&&&&&1&$8.8\cdot10^{ 3}$&  7 & 13 \\
&&&&&&2&$6.4\cdot10^{ 3}$&  7 & 15 \\
&&&&&&4&$3.3\cdot10^{ 3}$&  9 & 15 \\
&&&&&&7&$9.3\cdot10^{ 2}$& 17 & 22\rule[-1.5mm]{0mm}{3.mm}\\
 \hline
\end{tabular}
 \caption{Inclusive dijet
 cross section averaged over the regions indicated in $x$ and $Q^{2}$, for different values of $\Delta$
 as shown in Figures\,\ref{fig:stamp} and \ref{fig:dtmodnlo}. 
The measurement is restricted to values of the inelasticity variable
 $y$ between $0.1<y<0.7$ and to values of the polar angle of the scattered
 electron between $156^{\circ}< \theta < 175^{\circ}$. \label{tab:ddxsec}}
\end{table}
\newpage

\begin{table}[htp]
\footnotesize
\centering
 \begin{tabular}{|d{-1}cd{-1}|d{-1}cd{-1}|d{-1}|c|cc|}
 \hline
 \multicolumn{3}{|c|}
{\rule[-0.7mm]{0mm}{6.0mm} \footnotesize $Q^{2}$} & 
 \multicolumn{3}{c|}
{\footnotesize $x/10^{-4}$} & 
 \multicolumn{1}{c|}
{\footnotesize $\Delta$} & 
 \multicolumn{1}{c|}
{\footnotesize $\frac{d^{2}\sigma}{dQ^{2}dx}$} & 
 \multicolumn{1}{c}
 {\footnotesize $\delta_{stat}$} & 
 \multicolumn{1}{c|}
 {\footnotesize $\delta_{syst.}$} \\ 
 \multicolumn{3}{|c|}
 {$\rm\ [GeV^{2}]$} &
 \multicolumn{3}{c|} {} &
 \multicolumn{1}{c|} {$\rm\ [GeV]$} &
 \multicolumn{1}{c|} {$\rm\ [pb/GeV^{2}]$} &
 \multicolumn{1}{c} {$[\%]$} & 
 \multicolumn{1}{c|} {$[\%]$} 
 \rule[-2mm]{0mm}{6mm}\\
\hline
\rule[-0.1mm]{0mm}{3mm} 15.0&--& 20.0&  3.0&--& 5.0&0&$7.8\cdot10^{4}$&  6 &  9 \\
&&&&&&1&$7.6\cdot10^{ 4}$&  4 &  9 \\
&&&&&&2&$6.4\cdot10^{ 4}$&  4 &  9 \\
&&&&&&4&$3.7\cdot10^{ 4}$&  5 &  9 \\
&&&&&&7&$1.5\cdot10^{ 4}$&  7 & 10\rule[-1.5mm]{0mm}{3.mm}\\
 \hline
&&& \rule[-0.1mm]{0mm}{3mm}  5.0&--& 10.0&0&$4.5\cdot10^{ 4}$&  4& 11 \\
&&&&&&1&$4.0\cdot10^{ 4}$&  4 & 12 \\
&&&&&&2&$3.1\cdot10^{ 4}$&  4 & 11 \\
&&&&&&4&$1.6\cdot10^{ 4}$&  5 & 11 \\
&&&&&&7&$5.8\cdot10^{ 3}$&  8 & 12\rule[-1.5mm]{0mm}{3.mm}\\
 \hline
&&& \rule[-0.1mm]{0mm}{3mm} 10.0&--& 22.0&0&$1.1\cdot10^{ 4}$&  5 & 12 \\
&&&&&&1&$9.9\cdot10^{ 3}$&  5 & 13 \\
&&&&&&2&$7.6\cdot10^{ 3}$&  5 & 12 \\
&&&&&&4&$3.8\cdot10^{ 3}$&  7 & 13 \\
&&&&&&7&$1.3\cdot10^{ 3}$& 11 & 15\rule[-1.5mm]{0mm}{3.mm}\\
  \hline
 \rule[-0.1mm]{0mm}{3mm} 20.0&--& 30.0&  3.0&--&  5.0&0&$2.5\cdot10^{4}$&  5&  8 \\
&&&&&&1&$2.4\cdot10^{ 4}$&  5 &  8 \\
&&&&&&2&$2.1\cdot10^{ 4}$&  5 &  9 \\
&&&&&&4&$1.4\cdot10^{ 4}$&  6 &  8 \\
&&&&&&7&$6.7\cdot10^{ 3}$&  9 &  9\rule[-1.5mm]{0mm}{3.mm}\\
 \hline
&&& \rule[-0.1mm]{0mm}{3mm}  5.0&--& 10.0&0&$2.9\cdot10^{ 4}$&  3 & 9 \\
&&&&&&1&$2.7\cdot10^{ 4}$&  3 &   9 \\
&&&&&&2&$2.1\cdot10^{ 4}$&  3 &  10 \\
&&&&&&4&$1.2\cdot10^{ 4}$&  4 &  10 \\
&&&&&&7&$5.1\cdot10^{ 3}$&  6 &  10\rule[-1.5mm]{0mm}{3.mm}\\
 \hline
&&& \rule[-0.1mm]{0mm}{3mm}  10.0&--& 33.0&0&$6.2\cdot10^{ 3}$&  3 & 11 \\
&&&&&&1&$5.5\cdot10^{ 3}$&  3  &  11 \\
&&&&&&2&$4.1\cdot10^{ 3}$&  3  &  11 \\
&&&&&&4&$2.2\cdot10^{ 3}$&  4  &  12 \\
&&&&&&7&$9.3\cdot10^{ 2}$&  7  &  13\rule[-1.5mm]{0mm}{3.mm}\\
 \hline
\rule[-0.1mm]{0mm}{3mm} 30.0&--& 50.0&  5.0&--& 10.0&0&$1.0\cdot10^{4}$&  3 &  8 \\
&&&&&&1&$9.5\cdot10^{ 3}$&  3 &  8 \\
&&&&&&2&$8.2\cdot10^{ 3}$&  4 & 10 \\
&&&&&&4&$5.4\cdot10^{ 3}$&  4 &  9 \\
&&&&&&7&$2.6\cdot10^{ 3}$&  6 & 10\rule[-1.5mm]{0mm}{3.mm}\\
 \hline
&&& \rule[-0.1mm]{0mm}{3mm}  10.0&--& 55.0&0&$2.8\cdot10^{ 3}$&  2 & 10 \\
&&&&&&1&$2.5\cdot10^{ 3}$&  2 &  10 \\
&&&&&&2&$2.0\cdot10^{ 3}$&  3 &  11 \\
&&&&&&4&$1.1\cdot10^{ 3}$&  3 &  12 \\
&&&&&&7&$4.7\cdot10^{ 2}$&  5 &  13\rule[-1.5mm]{0mm}{3.mm}\\
 \hline
\rule[-0.1mm]{0mm}{3mm} 50.0&--&100.0& 10.0&--& 25.0&0&$1.4\cdot10^{3}$&  3 &  9 \\
&&&&&&1&$1.3\cdot10^{ 3}$&  3 &  9 \\
&&&&&&2&$1.1\cdot10^{ 3}$&  3 &  9 \\
&&&&&&4&$6.5\cdot10^{ 2}$&  4 & 10 \\
&&&&&&7&$2.8\cdot10^{ 2}$&  6 & 13\rule[-1.5mm]{0mm}{3.mm}\\
 \hline
&&& \rule[-0.1mm]{0mm}{3mm} 25.0&--&100.0&0&$4.5\cdot10^{ 2}$&  3 & 9 \\
&&&&&&1&$4.0\cdot10^{ 2}$&  3 &  10 \\
&&&&&&2&$3.2\cdot10^{ 2}$&  3 &  11 \\
&&&&&&4&$1.8\cdot10^{ 2}$&  4 &  11 \\
&&&&&&7&$7.1\cdot10^{ 1}$&  6 &  17\rule[-1.5mm]{0mm}{3.mm}\\
 \hline
\end{tabular}\\
\vspace*{\abovecaptionskip}
\normalsize
Table~\ref{tab:ddxsec} continued. 
\vspace*{\belowcaptionskip}
\end{table}

%% file: xstablept.tex
 \begin{table}[htb]
 \centering \footnotesize
 \begin{tabular}
{|d{-1}cd{-1}|d{-1}cd{-1}|d{-1}cd{-1}|c|cc|}
 \hline
 \multicolumn{3}{|c|}
{\rule[-0.7mm]{0mm}{6.0mm} $Q^{2}$} & 
 \multicolumn{3}{c|}
{$x/10^{-4}$} & 
 \multicolumn{3}{c|}
{$E^{\ast}_{T,max}$} & 
 \multicolumn{1}{c|}
{\normalsize $\frac{d^{3}\sigma}{dQ^{2}dx  dE^{\ast}_{T,max}}$} & 
 \multicolumn{1}{c}
 {$\delta_{stat}$} & 
 \multicolumn{1}{c|}
 {$\delta_{syst.}$} \\ 
 \multicolumn{3}{|c|}
 {\scriptsize $\rm\ [GeV^{2}]$} &
 \multicolumn{3}{c|} {} &
 \multicolumn{3}{c|} {\scriptsize $\rm\ [GeV]$} &
 \multicolumn{1}{c|} {\scriptsize $\rm\ [pb/GeV^{3}]$} &
 \multicolumn{1}{c} {\scriptsize $[\%]$} & 
 \multicolumn{1}{c|} {\scriptsize $[\%]$} 
 \rule[-2mm]{0mm}{6mm}\\
 \hline
\rule[-0.1mm]{0mm}{5mm} 5.0&--& 10.0&  1.0&--&  1.7&  7.0&--& 12.0&$6.8\cdot10^{4\phantom{-}}$ &  4&
 9 \\
&&&&&& 12.0&--& 20.0&$1.2\cdot10^{ 4\phantom{-}}$&  6 & 12 \\
&&&&&& 20.0&--& 30.0&$1.3\cdot10^{ 3\phantom{-}}$& 13 & 16 \\
&&&&&& 30.0&--& 60.0&$7.2\cdot10^{ 1\phantom{-}}$& 27 & 27\rule[-3mm]{0mm}{5mm}\\
 \hline
&&&  \rule[-0.1mm]{0mm}{5mm} 1.7&--&  3.0&  7.0&--& 12.0&$5.5\cdot10^{4\phantom{-}}$&  3 & 11 \\
&&&&&& 12.0&--& 20.0&$6.8\cdot10^{ 3\phantom{-}}$&  5 & 14 \\
&&&&&& 20.0&--& 30.0&$6.9\cdot10^{ 2\phantom{-}}$& 13 & 17 \\
&&&&&& 30.0&--& 60.0&$3.0\cdot10^{ 1\phantom{-}}$& 35 & 25\rule[-3mm]{0mm}{5mm}\\
 \hline
&&&  \rule[-0.1mm]{0mm}{5mm} 3.0&--&  5.0&  7.0&--& 12.0&$2.5\cdot10^{4\phantom{-}}$&  4 & 12 \\
&&&&&& 12.0&--& 20.0&$3.1\cdot10^{ 3\phantom{-}}$&  8 & 17 \\
&&&&&& 20.0&--& 30.0&$2.9\cdot10^{ 2\phantom{-}}$& 21 & 21\rule[-3mm]{0mm}{5mm}\\
 \hline
&&&  \rule[-0.1mm]{0mm}{5mm} 5.0&--& 10.0&  7.0&--& 12.0&$5.6\cdot10^{3\phantom{-}}$&  5 & 15 \\
&&&&&& 12.0&--& 20.0&$6.0\cdot10^{ 2\phantom{-}}$& 10 & 17 \\
&&&&&& 20.0&--& 30.0&$3.0\cdot10^{ 1\phantom{-}}$& 33 & 30\rule[-3mm]{0mm}{5mm}\\
 \hline
\rule[-0.1mm]{0mm}{5mm} 10.0&--& 30.0&  1.7&--&  3.0&  7.0&--& 12.0&$5.5\cdot10^{3\phantom{-}}$&  6 &  11 \\
&&&&&& 12.0&--& 20.0&$1.1 \cdot10^{ 3\phantom{-}}$&  7 & 12 \\
&&&&&& 20.0&--& 30.0&$1.0 \cdot10^{ 2\phantom{-}}$& 18 & 14 \\
&&&&&& 30.0&--& 60.0&$5.4 \cdot10^{ 0\phantom{-}}$& 36 & 22\rule[-3mm]{0mm}{5mm}\\
 \hline
&&& \rule[-0.1mm]{0mm}{5mm} 3.0&--&  5.0&  7.0&--& 12.0&$7.4 \cdot10^{3\phantom{-}}$&  3 & 11 \\
&&&&&& 12.0&--& 20.0&$1.2 \cdot10^{ 3\phantom{-}}$&  5 & 13 \\
&&&&&& 20.0&--& 30.0&$1.5 \cdot10^{ 2\phantom{-}}$& 11 & 18 \\
&&&&&& 30.0&--& 60.0&$1.0 \cdot10^{ 1\phantom{-}}$& 20 & 25\rule[-3mm]{0mm}{5mm}\\
 \hline
&&&  \rule[-0.1mm]{0mm}{5mm} 5.0&--&  1.0&  7.0&--& 12.0&$4.6 \cdot10^{3\phantom{-}}$&  2 & 12 \\
&&&&&& 12.0&--& 20.0&$6.2 \cdot10^{ 2\phantom{-}}$&  4 & 15 \\
&&&&&& 20.0&--& 30.0&$6.1 \cdot10^{ 1\phantom{-}}$& 11 & 18 \\
&&&&&& 30.0&--& 60.0&$1.4 \cdot10^{ 0\phantom{-}}$& 36 & 24\rule[-3mm]{0mm}{5mm}\\
 \hline
&&&  \rule[-0.1mm]{0mm}{5mm} 1.0&--& 33.0&  7.0&--& 12.0&$5.7\cdot10^{2\phantom{-}}$&  3 & 14 \\
&&&&&& 12.0&--& 20.0&$7.8 \cdot10^{ 1\phantom{-}}$&  6 & 16 \\
&&&&&& 20.0&--& 30.0&$4.6 \cdot10^{ 0\phantom{-}}$& 18 & 26 \\
&&&&&& 30.0&--& 60.0&$0.1 \cdot10^{ 0\phantom{-}}$& 69 & 23\rule[-3mm]{0mm}{5mm}\\
 \hline
\rule[-0.1mm]{0mm}{5mm} 30.0&--&100.0&  5.0&--& 10.0&  7.0&--& 12.0&$3.3\cdot10^{2\phantom{-}}$&  4 & 11 \\
&&&&&& 12.0&--& 20.0&$7.8\cdot10^{ 1\phantom{-}}$&  6 & 14 \\
&&&&&& 20.0&--& 30.0&$9.7\cdot10^{ 0\phantom{-}}$& 17 & 24 \\
&&&&&& 30.0&--& 60.0&$0.3\cdot10^{ 0\phantom{-}}$& 39 & 32\rule[-3mm]{0mm}{5mm}\\
 \hline
&&& \rule[-0.1mm]{0mm}{5mm} 10.0&--&100.0&  7.0&--& 12.0&$9.3\cdot10^{1\phantom{-}}$&  2 & 11 \\
&&&&&& 12.0&--& 20.0&$1.5 \cdot10^{ 1\phantom{-}}$&  3 & 17 \\
&&&&&& 20.0&--& 30.0&$1.3 \cdot10^{ 0\phantom{-}}$&  9 & 18 \\
&&&&&& 30.0&--& 60.0&$5.0 \cdot10^{-2}$& 26 & 30\rule[-3mm]{0mm}{5mm}\\
 \hline
 \end{tabular} \rm \normalsize
 \caption{Inclusive dijet
 cross section   
 averaged over the regions indicated in $x$, $Q^2$ and
 $E^{\ast}_{T,max}$ for $\Delta=2$~GeV as shown in Figure\,\ref{fig:ptjet}.
 The measurement is restricted to values of the inelasticity variable
 $y$ between \mbox{$0.1<y<0.7$} and to values of the polar angle of the scattered
 electron between \mbox{$156^{\circ}< \theta < 175^{\circ}$}.
}
 \label{tab:xsecpt}
 \end{table}

%% file: xstableeta.tex
 \begin{table}[htb]
 \centering \footnotesize
 \begin{tabular}
 {|d{-1}cd{-1}|d{-1}cd{-1}|d{-1}cd{-1}|c|cc|}
 \hline
 \multicolumn{3}{|c|}
{\rule[-0.7mm]{0mm}{6.0mm} \footnotesize $Q^{2}$} & 
 \multicolumn{3}{c|}
{\footnotesize $x/10^{-4}$} & 
 \multicolumn{3}{c|}
{\footnotesize $|\Delta\eta^{\ast}|$} & 
 \multicolumn{1}{c|}
{\normalsize $\frac{d^{3}\sigma}{dQ^{2}dx  d|\Delta\eta^{\ast}|}$} & 
 \multicolumn{1}{c}
 {\footnotesize $\delta_{stat}$} & 
 \multicolumn{1}{c|}
 {\footnotesize $\delta_{syst.}$} \\
 \multicolumn{3}{|c|}
 {$\rm\ [GeV^{2}]$} &
 \multicolumn{3}{c|} {} &
 \multicolumn{3}{c|} {} &
 \multicolumn{1}{c|} {$\rm\ [pb/GeV^{2}]$} &
 \multicolumn{1}{c} {$[\%]$} & 
 \multicolumn{1}{c|} {$[\%]$} 
 \rule[-2mm]{0mm}{6mm}\\
 \hline
\rule[-0.1mm]{0mm}{3.5mm} 5.0&--& 10.0&  1.0&--&  1.7&  0.0&--& 0.7&$2.3\cdot10^{ 5}$&  4 & 10 \\
&&&&&&  0.7&--&  1.4&$1.9\cdot10^{ 5}$&  5 & 11 \\
&&&&&&  1.4&--&  2.1&$1.3\cdot10^{ 5}$&  6 &  9 \\
&&&&&&  2.1&--&  2.8&$7.0\cdot10^{ 4}$&  9 &  9 \\
&&&&&&  2.8&--&  3.5&$2.0\cdot10^{ 4}$& 18 & 13\rule[-1.5mm]{0mm}{3mm}\\
 \hline
&&& \rule[-0.1mm]{0mm}{3.5mm} 1.7&--&  3.0&  0.0&--&
 0.7&$1.7\cdot10^{5}$&  4 & 12 \\
&&&&&&  0.7&--&  1.4&$1.4\cdot10^{ 5}$&  4 & 12 \\
&&&&&&  1.4&--&  2.1&$1.0\cdot10^{ 5}$&  6 & 14 \\
&&&&&&  2.1&--&  2.8&$5.1\cdot10^{ 4}$&  8 & 12 \\
&&&&&&  2.8&--&  3.5&$1.1\cdot10^{ 4}$& 17 & 30\rule[-1.5mm]{0mm}{3mm}\\
 \hline
&&& \rule[-0.1mm]{0mm}{3.5mm} 3.0&--&  5.0&  0.0&--&  0.7&$8.5\cdot10^{4}$&  5 & 14 \\
&&&&&&  0.7&--&  1.4&$6.9\cdot10^{ 4}$&  6 & 15 \\
&&&&&&  1.4&--&  2.1&$4.6\cdot10^{ 4}$&  8 & 13 \\
&&&&&&  2.1&--&  2.8&$1.7\cdot10^{ 4}$& 13 & 17\rule[-1.5mm]{0mm}{3mm}\\
 \hline
&&& \rule[-0.1mm]{0mm}{3.5mm} 5.0&--& 10.0&  0.0&--&  0.7&$2.0\cdot10^{4}$&  7 & 14 \\
&&&&&&  0.7&--&  1.4&$1.7\cdot10^{ 4}$&  8 & 16 \\
&&&&&&  1.4&--&  2.1&$8.6\cdot10^{ 3}$& 12 & 22\rule[-1.5mm]{0mm}{3mm}\\
 \hline
\rule[-0.1mm]{0mm}{3.5mm} 10.0&--& 30.0&  1.7&--&  3.0&  0.0&--&0.7&$2.0\cdot10^{ 4}$&  5 & 13 \\
&&&&&&  0.7&--&  1.4&$1.5\cdot10^{ 4}$&  6 &  10 \\
&&&&&&  1.4&--&  2.1&$1.2\cdot10^{ 4}$&  7 &  10 \\
&&&&&&  2.1&--&  2.8&$5.0\cdot10^{ 3}$& 11 &  13 \\
&&&&&&  2.8&--&  3.5&$1.8\cdot10^{ 3}$& 20 &  31\rule[-1.5mm]{0mm}{3mm}\\
 \hline
&&& \rule[-0.1mm]{0mm}{3.5mm} 3.0&--&  5.0&  0.0&--&  0.7&$2.5\cdot10^{4}$&  4 & 12 \\
&&&&&&  0.7&--&  1.4&$1.9\cdot10^{ 4}$&  4 & 11 \\
&&&&&&  1.4&--&  2.1&$1.4\cdot10^{ 4}$&  5 & 11 \\
&&&&&&  2.1&--&  2.8&$9.4\cdot10^{ 3}$&  7 & 11 \\
&&&&&&  2.8&--&  3.5&$2.0\cdot10^{ 3}$& 13 & 13\rule[-1.5mm]{0mm}{3mm}\\
 \hline
&&& \rule[-0.1mm]{0mm}{3.5mm}   5.0&--&  1.0&  0.0&--& 0.7&$1.5\cdot10^{ 4}$&  3 & 12 \\
&&&&&&  0.7&--&  1.4&$1.2\cdot10^{ 4}$&  3 & 12 \\
&&&&&&  1.4&--&  2.1&$8.1\cdot10^{ 3}$&  5 & 15 \\
&&&&&&  2.1&--&  2.8&$4.0\cdot10^{ 3}$&  7 & 14 \\
&&&&&&  2.8&--&  3.5&$8.8\cdot10^{ 2}$& 16 & 30\rule[-1.5mm]{0mm}{3mm}\\
 \hline
&&& \rule[-0.1mm]{0mm}{3.5mm} 1.0&--& 33.0&  0.0&--& 0.7&$2.1\cdot10^{3}$&  4 & 13 \\
&&&&&&  0.7&--&  1.4&$1.8\cdot10^{ 3}$&  4 & 15 \\
&&&&&&  1.4&--&  2.1&$9.5\cdot10^{ 2}$&  7 & 16 \\
&&&&&&  2.1&--&  2.8&$2.2\cdot10^{ 2}$& 13 & 26\rule[-1.5mm]{0mm}{3mm}\\
 \hline
\rule[-0.1mm]{0mm}{3.5mm} 30.0&--&100.0&  5.0&--& 10.0&  0.0&--& 0.7&$1.1\cdot10^{ 3}$&  6 & 11 \\
&&&&&&  0.7&--&  1.4&$1.1\cdot10^{ 3}$&  6 & 14 \\
&&&&&&  1.4&--&  2.1&$6.8\cdot10^{ 2}$&  7 & 12 \\
&&&&&&  2.1&--&  2.8&$3.4\cdot10^{ 2}$& 11 & 13 \\
&&&&&&  2.8&--&  3.5&$1.6\cdot10^{ 2}$& 21 & 24\rule[-1.5mm]{0mm}{3mm}\\
 \hline
&&& \rule[-0.1mm]{0mm}{3.5mm} 10.0&--&100.0&  0.0&--&  0.7&$3.4\cdot10^{
2}$&  3 & 14 \\
&&&&&&  0.7&--&  1.4&$2.7\cdot10^{ 2}$&  3 & 11 \\
&&&&&&  1.4&--&  2.1&$1.8\cdot10^{ 2}$&  4 & 13 \\
&&&&&&  2.1&--&  2.8&$6.1\cdot10^{ 1}$&  6 & 14 \\
&&&&&&  2.8&--&  3.5&$9.1\cdot10^{ 0}$& 15 & 21\rule[-1.5mm]{0mm}{3mm}\\
 \hline
 \end{tabular} \rm \normalsize
 \caption{Inclusive dijet
 cross section  averaged over the regions indicated in $x$, $Q^2$ and $|\Delta\eta^{\ast}|$ for $\Delta=2$~GeV
as shown in Figure\,\ref{fig:etajet}.
 The measurement is restricted to values of the inelasticity variable
 $y$ between \mbox{$0.1<y<0.7$} and to values of the polar angle of the scattered
 electron between \mbox{$156^{\circ}< \theta < 175^{\circ}$}.
 \label{tab:xseceta}}
 \end{table}

%% file: r2dtablept.tex
\begin{table}[htp]
 \centering
 \footnotesize
 \begin{tabular}{|d{-1}cd{-1}|d{-1}cd{-1}|d{-1}|c|cc|}
 \hline
  \multicolumn{3}{|c|}
{\rule[-0.7mm]{0mm}{6.0mm} \footnotesize $Q^{2}$} & 
 \multicolumn{3}{c|}
{\footnotesize $x/10^{-4}$} & 
 \multicolumn{1}{c|}
{\footnotesize $\Delta$} & 
 \multicolumn{1}{c|}
{\normalsize $R_{2}=\frac{N_{dijet}}{N_{DIS}} $} & 
 \multicolumn{1}{c}
 {\footnotesize $\delta_{stat}$} & 
 \multicolumn{1}{c|}
 {\footnotesize $\delta_{syst.}$} \\
 \multicolumn{3}{|c|}
 {$\rm\ [GeV^{2}]$} &
 \multicolumn{3}{c|} {} &
\multicolumn{1}{c|} {$\rm\ [GeV]$} &
 \multicolumn{1}{c|} {} &
 \multicolumn{1}{c} {$[\%]$} & 
 \multicolumn{1}{c|} {$[\%]$} 
 \rule[-2mm]{0mm}{6mm}\\
 \hline
 \rule[-0.1mm]{0mm}{3mm}  5.0&--& 10.0&  1.0&--&  1.7&0&  0.031&  4 & 9 \\
&&&&&&1&  0.028&  4 &  9 \\
&&&&&&2&  0.023&  3 &  9 \\
&&&&&&4&  0.014&  4 &  9 \\
&&&&&&7&  0.006&  6 &  10\rule[-1.5mm]{0mm}{3.mm}\\
 \hline
&&& \rule[-0.1mm]{0mm}{3mm} 1.7&--&  3.0&0&  0.027&  3 &  9\\
&&&&&&1&  0.024&  3 &  9 \\
&&&&&&2&  0.019&  3 &  9 \\
&&&&&&4&  0.010&  4 &  10 \\
&&&&&&7&  0.004&  5 &  12\rule[-1.5mm]{0mm}{3.mm}\\
 \hline
&&&   \rule[-0.1mm]{0mm}{3mm} 3.0&--&  5.0&0&  0.021&  4& 12 \\
&&&&&&1&  0.019&  4& 12 \\
&&&&&&2&  0.014&  4& 11 \\
&&&&&&4&  0.007&  5& 11 \\
&&&&&&7&  0.003&  8& 13\rule[-1.5mm]{0mm}{3.mm}\\
 \hline
&&&  \rule[-0.1mm]{0mm}{3mm} 5.0&--& 10.0&0&  0.015&  5& 15 \\
&&&&&&1&  0.013&  5 & 14 \\
&&&&&&2&  0.010&  5 & 14 \\
&&&&&&4&  0.004&  7 & 14 \\
&&&&&&7&  0.002& 10 & 18\rule[-1.5mm]{0mm}{3.mm}\\
 \hline
\rule[-0.1mm]{0mm}{3mm} 10.0&--& 15.0&  1.7&--&  3.0&0&  0.040&  4&  8\\
&&&&&&1&  0.037&  4&  8\\
&&&&&&2&  0.031&  4&  9\\
&&&&&&4&  0.020&  5& 10\\
&&&&&&7&  0.009&  7&  9\rule[-1.5mm]{0mm}{3.mm}\\
 \hline
&&&   \rule[-0.1mm]{0mm}{3mm} 3.0&--&  5.0&0&  0.035&  3&  10\\
&&&&&&1&  0.031&  3&  9\\
&&&&&&2&  0.024&  4&  9\\
&&&&&&4&  0.013&  5& 10\\
&&&&&&7&  0.005&  7& 11\rule[-1.5mm]{0mm}{3.mm}\\
 \hline
&&&  \rule[-0.1mm]{0mm}{3mm} 5.0&--& 10.0&0&  0.027&  3& 13\\
&&&&&&1&  0.024&  4& 12\\
&&&&&&2&  0.018&  4& 13\\
&&&&&&4&  0.009&  5& 13\\
&&&&&&7&  0.003&  7& 14\rule[-1.5mm]{0mm}{3.mm}\\
 \hline
&&& \rule[-0.1mm]{0mm}{3mm} 10.0&--& 18.0&0&  0.019&  6& 15\\
&&&&&&1&  0.016&  7& 14\\
&&&&&&2&  0.011&  7& 18\\
&&&&&&4&  0.006&  9& 20\\
&&&&&&7&  0.002& 17& 24\rule[-1.5mm]{0mm}{3.mm}\\
 \hline
\end{tabular}
 \caption{
 Dijet rate $R_2$ averaged over the regions indicated in $\qsq$ and $\myx$ for different values of $\Delta$.
 The measurement is restricted to values of the inelasticity variable
 $y$ between $0.1<y<0.7$ and to values of the polar angle of the scattered
 electron between $156^{\circ}< \theta < 175^{\circ}$.
 \label{tab:r2sa}
}
\end{table}
\newpage

\begin{table}[htp]
\centering
\footnotesize
 \begin{tabular}{|d{-1}cd{-1}|d{-1}cd{-1}|d{-1}|c|cc|}
 \hline
 \multicolumn{3}{|c|}
{\rule[-0.7mm]{0mm}{6.0mm} \footnotesize $Q^{2}$} & 
 \multicolumn{3}{c|}
{\footnotesize $x/10^{-4}$} & 
 \multicolumn{1}{c|}
{\footnotesize $\Delta$} & 
 \multicolumn{1}{c|}
{\footnotesize $R_{2}=\frac{N_{dijet}}{N_{DIS}}$} & 
 \multicolumn{1}{c}
 {\footnotesize $\delta_{stat}$} & 
 \multicolumn{1}{c|}
 {\footnotesize $\delta_{syst.}$} \\ 
 \multicolumn{3}{|c|}
 {$\rm\ [GeV^{2}]$} &
 \multicolumn{3}{c|} {} &
 \multicolumn{1}{c|} {$\rm\ [GeV]$} &
 \multicolumn{1}{c|} {} &
 \multicolumn{1}{c} {$[\%]$} & 
 \multicolumn{1}{c|} {$[\%]$} 
 \rule[-2mm]{0mm}{6mm}\\
\hline
\rule[-0.1mm]{0mm}{3mm} 15.0&--& 20.0&  3.0&--&  5.0&0&  0.043&  6&  9\\
&&&&&&1&  0.042&  4&  8\\
&&&&&&2&  0.035&  4&  9\\
&&&&&&4&  0.020&  5&  9\\
&&&&&&7&  0.008&  7& 10\rule[-1.5mm]{0mm}{3.mm}\\
 \hline
&&&  \rule[-0.1mm]{0mm}{3mm} 5.0&--& 10.0&0&  0.041&  4& 10\\
&&&&&&1&  0.036&  4& 12\\
&&&&&&2&  0.028&  4& 11\\
&&&&&&4&  0.015&  5& 10\\
&&&&&&7&  0.005&  8& 12\rule[-1.5mm]{0mm}{3.mm}\\
 \hline
&&& \rule[-0.1mm]{0mm}{3mm} 10.0&--& 22.0&0&  0.027&  5& 13\\
&&&&&&1&  0.024&  5& 13\\
&&&&&&2&  0.019&  5& 12\\
&&&&&&4&  0.009&  7& 14\\
&&&&&&7&  0.003& 11& 17\rule[-1.5mm]{0mm}{3.mm}\\
 \hline
\rule[-0.1mm]{0mm}{3mm} 20.0&--& 30.0&  3.0&--&  5.0&0&  0.056&  5&  8\\
&&&&&&1&  0.052&  5&  8\\
&&&&&&2&  0.046&  6&  8\\
&&&&&&4&  0.030&  7&  9\\
&&&&&&7&  0.015&  9& 10\rule[-1.5mm]{0mm}{3.mm}\\
 \hline
&&&  \rule[-0.1mm]{0mm}{3mm} 5.0&--& 10.0&0&  0.054&  3&  9\\
&&&&&&1&  0.049&  3&  9\\
&&&&&&2&  0.040&  3&  9\\
&&&&&&4&  0.023&  4& 10\\
&&&&&&7&  0.010&  6& 10\rule[-1.5mm]{0mm}{3.mm} \\
 \hline
&&& \rule[-0.1mm]{0mm}{3mm} 10.0&--& 33.0&0&  0.038&  3& 11\\
&&&&&&1&  0.033&  3& 11\\
&&&&&&2&  0.025&  4& 12\\
&&&&&&4&  0.014&  4& 13\\
&&&&&&7&  0.006&  7& 15\rule[-1.5mm]{0mm}{3.mm}\\
 \hline
\rule[-0.1mm]{0mm}{3mm} 30.0&--& 50.0&  5.0&--& 10.0&0&  0.068&  3&  8\\
&&&&&&1&  0.064&  3&  8\\
&&&&&&2&  0.056&  4&  9\\
&&&&&&4&  0.036&  4&  9\\
&&&&&&7&  0.018&  6&  10\rule[-1.5mm]{0mm}{3.mm}\\
 \hline
&&&  \rule[-0.1mm]{0mm}{3mm} 10.0&--& 55.0&0&  0.056&  2&  10\\
&&&&&&1&  0.051&  2& 10\\
&&&&&&2&  0.041&  3& 11\\
&&&&&&4&  0.023&  3& 12\\
&&&&&&7&  0.009&  5& 13\rule[-1.5mm]{0mm}{3.mm}\\
 \hline
\rule[-0.1mm]{0mm}{3mm} 50.0&--&100.0& 10.0&--& 25.0&0&  0.087&  3&  9\\
&&&&&&1&  0.080&  3&  8\\
&&&&&&2&  0.067&  3&  9\\
&&&&&&4&  0.041&  4& 10\\
&&&&&&7&  0.018&  6& 13\rule[-1.5mm]{0mm}{3.mm}\\
 \hline
&&& \rule[-0.1mm]{0mm}{3mm} 25.0&--&100.0&0&  0.072&  3&  9\\
&&&&&&1&  0.065&  3& 10\\
&&&&&&2&  0.052&  3& 11\\
&&&&&&4&  0.029&  4& 12\\
&&&&&&7&  0.011&  6& 18\rule[-1.5mm]{0mm}{3.mm}\\
 \hline
\end{tabular}\\
\vspace*{\abovecaptionskip}
\normalsize
Table~\ref{tab:r2sa} continued. 
\vspace*{\belowcaptionskip}
\end{table}

%% file: rstablept.tex
 \begin{table}[htb]
 \centering \footnotesize 
 \begin{tabular}
 {|d{-1}cd{-1}|d{-1}cd{-1}|d{-1}cd{-1}|c|cc|}
 \hline
 \multicolumn{3}{|c|}
{\rule[-0.7mm]{0mm}{6.0mm} $Q^{2}$} & 
 \multicolumn{3}{c|}
{$x/10^{-4}$} & 
 \multicolumn{3}{c|}
{$E^{\ast}_{T,max}$} & 
\multicolumn{1}{c|}
{\normalsize $\frac{dR_{2}}{dE^{\ast}_{T,max}}$} & 
 \multicolumn{1}{c}
 {$\delta_{stat}$} & 
 \multicolumn{1}{c|}
 {$\delta_{syst.}$} \\ 
 \multicolumn{3}{|c|}
 {\scriptsize $\rm\ [GeV^{2}]$} &
 \multicolumn{3}{c|} {} &
 \multicolumn{3}{c|} {\scriptsize $\rm\ [GeV]$} &
 \multicolumn{1}{c|} {\scriptsize $\rm\ [GeV^{-1}]$} &
 \multicolumn{1}{c} {\scriptsize $[\%]$} & 
 \multicolumn{1}{c|} {\scriptsize $[\%]$} 
 \rule[-2mm]{0mm}{6mm}\\
 \hline
\rule[-0.1mm]{0mm}{5mm}  5.0&--& 10.0&  1.0&--&  1.7&  7.0&--& 12.0&$3.5\cdot10^{-3}$&  3& 10\\
&&&&&& 12.0&--& 20.0&$4.9\cdot10^{-4}$&  6& 14\\
&&&&&& 20.0&--& 30.0&$6.8\cdot10^{-5}$& 13& 18\\
&&&&&& 30.0&--& 60.0&$3.6\cdot10^{-6}$& 27& 27\rule[-3mm]{0mm}{5mm}\\
 \hline
&&& \rule[-0.1mm]{0mm}{5mm} 1.7&--&  3.0&  7.0&--& 12.0&$3.1\cdot10^{-3}$&  3& 11\\
&&&&&& 12.0&--& 20.0&$3.1\cdot10^{-4}$&  5& 14\\
&&&&&& 20.0&--& 30.0&$4.0\cdot10^{-5}$& 13& 18\\
&&&&&& 30.0&--& 60.0&$1.7\cdot10^{-6}$& 35& 25\rule[-3mm]{0mm}{5mm}\\
 \hline
&&& \rule[-0.1mm]{0mm}{5mm} 3.0&--&  5.0&  7.0&--& 12.0&$2.3\cdot10^{-3}$&  4& 13\\
&&&&&& 12.0&--& 20.0&$2.3\cdot10^{-4}$&  8& 17\\
&&&&&& 20.0&--& 30.0&$2.6\cdot10^{-5}$& 21& 22\rule[-3mm]{0mm}{5mm}\\
 \hline
&&& \rule[-0.1mm]{0mm}{5mm} 5.0&--& 10.0&  7.0&--& 12.0&$1.6\cdot10^{-3}$&  5& 16\\
&&&&&& 12.0&--& 20.0&$1.4\cdot10^{-4}$& 10& 20\\
&&&&&& 20.0&--& 30.0&$7.8\cdot10^{-6}$& 34& 36\rule[-3mm]{0mm}{5mm}\\
 \hline
\rule[-0.1mm]{0mm}{5mm} 10.0&--& 30.0&  1.7&--&  3.0&  7.0&--& 12.0&$4.6\cdot10^{-3}$&  5& 11\\
&&&&&& 12.0&--& 20.0&$7.2\cdot10^{-4}$&  7& 13\\
&&&&&& 20.0&--& 30.0&$8.6\cdot10^{-5}$& 18& 15\\
&&&&&& 30.0&--& 60.0&$4.5\cdot10^{-6}$& 36& 23\rule[-3mm]{0mm}{5mm}\\
 \hline
&&&  \rule[-0.1mm]{0mm}{5mm} 3.0&--&  5.0&  7.0&--& 12.0&$4.6\cdot10^{-3}$&  3& 11\\
&&&&&& 12.0&--& 20.0&$5.9\cdot10^{-4}$&  5& 13\\
&&&&&& 20.0&--& 30.0&$9.2\cdot10^{-5}$& 11& 18\\
&&&&&& 30.0&--& 60.0&$6.4\cdot10^{-6}$& 20& 25\rule[-3mm]{0mm}{5mm}\\
 \hline
&&& \rule[-0.1mm]{0mm}{5mm} 5.0&--&  1.0&  7.0&--& 12.0&$4.2\cdot10^{-3}$&  2& 12\\
&&&&&& 12.0&--& 20.0&$4.6\cdot10^{-4}$&  4& 15\\
&&&&&& 20.0&--& 30.0&$5.6\cdot10^{-5}$& 11& 19\\
&&&&&& 30.0&--& 60.0&$1.3\cdot10^{-6}$& 36& 25\rule[-3mm]{0mm}{5mm}\\
 \hline
&&& \rule[-0.1mm]{0mm}{5mm} 1.0&--& 33.0&  7.0&--& 12.0&$3.1\cdot10^{-3}$&  3& 14\\
&&&&&& 12.0&--& 20.0&$3.4\cdot10^{-4}$&  6& 17\\
&&&&&& 20.0&--& 30.0&$2.5\cdot10^{-5}$& 18& 27\\
&&&&&& 30.0&--& 60.0&$7.0\cdot10^{-7}$& 69& 25\rule[-3mm]{0mm}{5mm}\\
 \hline
\rule[-0.1mm]{0mm}{5mm} 30.0&--&100.0&  5.0&--& 10.0&  7.0&--& 12.0&$7.7\cdot10^{-3}$&  4& 11\\
&&&&&& 12.0&--& 20.0&$1.5\cdot10^{-3}$&  6& 13\\
&&&&&& 20.0&--& 30.0&$2.3\cdot10^{-4}$& 17& 23\\
&&&&&& 30.0&--& 60.0&$7.3\cdot10^{-6}$& 39& 32\rule[-3mm]{0mm}{5mm}\\
 \hline
&&& \rule[-0.1mm]{0mm}{5mm} 10.0&--&100.0&  7.0&--& 12.0&$7.4\cdot10^{-3}$&  2& 11\\
&&&&&& 12.0&--& 20.0&$9.5\cdot10^{-4}$&  3& 17\\
&&&&&& 20.0&--& 30.0&$1.1\cdot10^{-4}$&  9& 19\\
&&&&&& 30.0&--& 60.0&$4.0\cdot10^{-6}$& 26& 31\rule[-3mm]{0mm}{5mm}\\
 \hline
 \end{tabular} \rm \normalsize
 \caption{
 Dijet rate $R_2$ averaged over the regions indicated in $x$, $Q^2$
 and $E^{\ast}_{T,max}$ for $\Delta = 2$~GeV.
The measurement is restricted to values of the inelasticity variable
 $y$ between $0.1<y<0.7$ and to values of the polar angle of the scattered
 electron between $156^{\circ}< \theta < 175^{\circ}$.
}
 \label{tab:r2pt}
 \end{table}

%% file: rstableeta.tex
  \begin{table}[htb]
 \centering \footnotesize 
 \begin{tabular}
 {|d{-1}cd{-1}|d{-1}cd{-1}|d{-1}cd{-1}|c|cc|}
  \hline
 \multicolumn{3}{|c|}
{\rule[-0.7mm]{0mm}{6.0mm} \footnotesize $Q^{2}$} & 
 \multicolumn{3}{c|}
{\footnotesize $x/10^{-4}$} & 
 \multicolumn{3}{c|}
{\footnotesize $|\Delta\eta^{\ast}|$} & 
 \multicolumn{1}{c|}
{\normalsize $\frac{dR_{2}}{d|\Delta\eta^{\ast}|}$} & 
 \multicolumn{1}{c}
 {\footnotesize $\delta_{stat}$} & 
 \multicolumn{1}{c|}
 {\footnotesize $\delta_{syst.}$} \\
 \multicolumn{3}{|c|}
 {$\rm\ [GeV^{2}]$} &
 \multicolumn{3}{c|} {} &
 \multicolumn{3}{c|} {} &
 \multicolumn{1}{c|} {} &
 \multicolumn{1}{c} {$[\%]$} & 
 \multicolumn{1}{c|} {$[\%]$} 
 \rule[-2mm]{0mm}{6mm}\\
 \hline
\rule[-0.1mm]{0mm}{3mm}  5.0&--& 10.0&  1.0&--&  1.7&  0.0&--&  0.7&$1.2\cdot10^{-2}$&  5& 12\\
&&&&&&  0.7&--&  1.4&$9.5\cdot10^{-3}$&  5& 12\\
&&&&&&  1.4&--&  2.1&$6.5\cdot10^{-3}$&  7& 11\\
&&&&&&  2.1&--&  2.8&$3.5\cdot10^{-3}$&  9& 10\\
&&&&&&  2.8&--&  3.5&$9.9\cdot10^{-4}$& 18& 13\rule[-1.5mm]{0mm}{3mm}\\
 \hline
&&&  \rule[-0.1mm]{0mm}{3mm} 1.7&--&  3.0&  0.0&--&  0.7&$9.6\cdot10^{-3}$&  4& 12\\
&&&&&&  0.7&--&  1.4&$8.0\cdot10^{-3}$&  5& 12\\
&&&&&&  1.4&--&  2.1&$5.8\cdot10^{-3}$&  6& 13\\
&&&&&&  2.1&--&  2.8&$2.9\cdot10^{-3}$&  8& 12\\
&&&&&&  2.8&--&  3.5&$6.3\cdot10^{-4}$& 17& 31\rule[-1.5mm]{0mm}{3mm}\\
 \hline
&&& \rule[-0.1mm]{0mm}{3mm} 3.0&--&  5.0&  0.0&--&  0.7&$7.7\cdot10^{-3}$&  5& 14\\
&&&&&&  0.7&--&  1.4&$6.3\cdot10^{-3}$&  6& 15\\
&&&&&&  1.4&--&  2.1&$4.2\cdot10^{-3}$&  8& 15\\
&&&&&&  2.1&--&  2.8&$1.5\cdot10^{-3}$& 13& 18\rule[-1.5mm]{0mm}{3mm}\\
 \hline
&&&  \rule[-0.1mm]{0mm}{3mm} 5.0&--& 10.0&  0.0&--&  0.7&$5.7\cdot10^{-3}$&  7& 14\\
&&&&&&  0.7&--&  1.4&$4.9\cdot10^{-3}$&  8& 17\\
&&&&&&  1.4&--&  2.1&$2.4\cdot10^{-3}$& 12& 25\rule[-1.5mm]{0mm}{3mm}\\
 \hline
\rule[-0.1mm]{0mm}{3mm} 10.0&--& 30.0&  1.7&--&  3.0&  0.0&--&  0.7&$1.7\cdot10^{-2}$&  5& 13\\
&&&&&&  0.7&--&  1.4&$1.3\cdot10^{-2}$&  6& 11\\
&&&&&&  1.4&--&  2.1&$9.8\cdot10^{-3}$&  7& 10\\
&&&&&&  2.1&--&  2.8&$4.2\cdot10^{-3}$& 11& 14\\
&&&&&&  2.8&--&  3.5&$1.5\cdot10^{-3}$& 20& 30\rule[-1.5mm]{0mm}{3mm}\\
 \hline
&&& \rule[-0.1mm]{0mm}{3mm} 3.0&--&  5.0&  0.0&--&  0.7&$1.5\cdot10^{-2}$&  4& 12\\
&&&&&&  0.7&--&  1.4&$1.2\cdot10^{-2}$&  4& 11\\
&&&&&&  1.4&--&  2.1&$8.8\cdot10^{-3}$&  5& 12\\
&&&&&&  2.1&--&  2.8&$5.8\cdot10^{-3}$&  7& 11\\
&&&&&&  2.8&--&  3.5&$1.2\cdot10^{-3}$& 13& 13\rule[-1.5mm]{0mm}{3mm}\\
 \hline
&&& \rule[-0.1mm]{0mm}{3mm}  5.0&--&  1.0&  0.0&--&  0.7&$1.4\cdot10^{-2}$&  3& 12\\
&&&&&&  0.7&--&  1.4&$1.1\cdot10^{-2}$&  3& 12\\
&&&&&&  1.4&--&  2.1&$7.5\cdot10^{-3}$&  5& 15\\
&&&&&&  2.1&--&  2.8&$3.7\cdot10^{-3}$&  7& 14\\
&&&&&&  2.8&--&  3.5&$8.1\cdot10^{-4}$& 16& 30\rule[-1.5mm]{0mm}{3mm}\\
 \hline
&&& \rule[-0.1mm]{0mm}{3mm} 1.0&--& 33.0&  0.0&--&  0.7&$1.2\cdot10^{-2}$&  4& 13\\
&&&&&&  0.7&--&  1.4&$9.6\cdot10^{-3}$&  4& 15\\
&&&&&&  1.4&--&  2.1&$5.2\cdot10^{-3}$&  7& 18\\
&&&&&&  2.1&--&  2.8&$1.2\cdot10^{-3}$& 13& 28\rule[-1.5mm]{0mm}{3mm}\\
 \hline
 \rule[-0.1mm]{0mm}{3mm} 30.0&--&100.0&  5.0&--& 10.0&  0.0&--&  0.7&$2.6\cdot10^{-2}$&  6& 11\\
&&&&&&  0.7&--&  1.4&$2.6\cdot10^{-2}$&  6& 14\\
&&&&&&  1.4&--&  2.1&$1.6\cdot10^{-2}$&  7& 12\\
&&&&&&  2.1&--&  2.8&$7.9\cdot10^{-3}$& 11& 13\\
&&&&&&  2.8&--&  3.5&$3.8\cdot10^{-3}$& 21& 24\rule[-1.5mm]{0mm}{3mm}\\
 \hline
&&& \rule[-0.1mm]{0mm}{3mm} 10.0&--&100.0&  0.0&--&  0.7&$2.7\cdot10^{-2}$&  3& 13\\
&&&&&&  0.7&--&  1.4&$2.2\cdot10^{-2}$&  3& 11\\
&&&&&&  1.4&--&  2.1&$1.4\cdot10^{-2}$&  4& 12\\
&&&&&&  2.1&--&  2.8&$4.9\cdot10^{-3}$&  6& 14\\
&&&&&&  2.8&--&  3.5&$7.3\cdot10^{-4}$& 15& 22\rule[-1.5mm]{0mm}{3mm}\\
 \hline
 \end{tabular} \rm \normalsize
 \caption{
Dijet rate $R_2$ averaged over the regions indicated in $x$, $Q^2$ and
$|\Delta\eta^{\ast}|$ for $\Delta = 2$~GeV.
The measurement is restricted to values of the inelasticity variable
$y$ between $0.1<y<0.7$ and to values of the polar angle of the scattered
electron between $156^{\circ}< \theta < 175^{\circ}$.
}
 \label{tab:r2eta}
 \end{table}

%% file: xstabless.tex
 \begin{table}[htb]
\centering \footnotesize 
 \begin{tabular}
 {|d{-1}cd{-1}|d{-1}cd{-1}|c|cc|}
 \hline
 \multicolumn{3}{|c|}
{\rule[-0.7mm]{0mm}{6.0mm} $Q^{2}$} & 
 \multicolumn{3}{c|}
{$x/10^{-4}$} & 
 \multicolumn{1}{c|}
 {$S$} & 
 \multicolumn{1}{c}
 {$\delta_{stat}$} & 
 \multicolumn{1}{c|}
 {$\delta_{syst.}$} \\ 
\multicolumn{3}{|c|}
 {\scriptsize $\rm\ [GeV^{2}]$} &
 \multicolumn{3}{c|} {} &
 \multicolumn{1}{c|} {} &
 \multicolumn{1}{c} {\scriptsize $[\%]$} & 
 \multicolumn{1}{c|} {\scriptsize $[\%]$} 
 \rule[-2mm]{0mm}{6mm}\\
 \hline
\rule[-0.1mm]{0mm}{5mm}  5.0&--& 10.0&  1.0&--&  1.7&  0.086 &  8 &  9 \\
&&&  1.7&--&  3.0&  0.053 &  9 &  9 \\
&&&  3.0&--&  5.0&  0.049 & 13 &  9 \\
&&&  5.0&--& 10.0&  0.038 & 21 & 11\rule[-3mm]{0mm}{5mm}\\
 \hline
\rule[-0.1mm]{0mm}{5mm} 10.0&--& 15.0&  1.7&--&  3.0&  0.056 & 11 &  9\\
&&&  3.0&--&  5.0&  0.048 & 13 & 12\\
&&&  5.0&--& 10.0&  0.039 & 15 &  10\\
&&& 10.0&--& 18.0&  0.022 & 33 & 24\rule[-3mm]{0mm}{5mm}\\
\hline
\rule[-0.1mm]{0mm}{5mm} 15.0&--& 20.0&  3.0&--&  5.0&  0.066 & 13 &  7\\
&&&  5.0&--& 10.0&  0.050 & 15 & 16\\
&&& 10.0&--& 22.0&  0.039 & 23 & 24\rule[-3mm]{0mm}{5mm}\\
 \hline
\rule[-0.1mm]{0mm}{5mm}  20.0&--& 30.0&  3.0&--&  5.0&  0.086 & 16 & 14\\
&&&  5.0&--& 10.0&  0.051 & 12 &  12\\
&&& 10.0&--& 33.0&  0.043 & 15 & 13\rule[-3mm]{0mm}{5mm}\\
 \hline
\rule[-0.1mm]{0mm}{5mm} 30.0&--& 50.0&  5.0&--& 10.0&  0.058& 13 & 10\\
&&& 10.0&--& 55.0&  0.038& 11 & 16\rule[-3mm]{0mm}{5mm}\\
 \hline
\rule[-0.1mm]{0mm}{5mm} 50.0&--&100.0& 10.0&--& 25.0&  0.040& 14 & 16\\
&&& 25.0&--&100.0&  0.038& 13 &  11\rule[-3mm]{0mm}{5mm}\\
 \hline
 \end{tabular} \rm \normalsize
 \caption{
Measured ratio $S$ for jets with an azimuthal separation
of $\Delta\phi^{\ast}<120^{\circ}$ for $\Delta = 2$~GeV as shown in 
Figures\,\ref{fig:ssjet120} to~\ref{fig:ssjet120casc}. 
The measurement is restricted to values of the inelasticity variable
$y$ between $0.1<y<0.7$ and to values of the polar angle of the scattered
electron between $156^{\circ}< \theta < 175^{\circ}$.
}
\label{tab:ssjets}
\end{table}